\documentclass[12pt]{article}

\usepackage{graphicx}

\title{Observations and perspectives on the diversification of genomes}
\author{Dirson Jian Li \footnote{E-Mail: dirson@mail.xjtu.edu.cn}\\ \normalsize{\it \small Department of Applied Physics, School of Science, Xi'an Jiaotong University, Xi'an 710049, China} }

\date{}
\begin{document}
\maketitle
\sloppy
\baselineskip18pt
\setlength{\parskip}{12pt}

{\bf \begin{center}{Abstract}\end{center}
Rich information on the prebiotic evolution is still stored in contemporary genomic data. The statistical mechanism at the sequence level may play a significant role in the prebiotic evolution. Based on statistical analysis of genome sequences, it has been observed that there is a close relationship between the evolution of the genetic code and the organisation of genomes. A biodiversity space for species is constructed based on comparing the distributions of codons in genomes for different species according to recruitment order of codons in the prebiotic evolution, by which a closely relationship between the evolution of the genetic code and the tree of life has been confirmed. On one hand, the three domain tree of life can be reconstructed according to the distance matrix of species in this biodiversity space, which supports the three-domain tree rather than the eocyte tree. On the other hand, an evolutionary tree of codons can be obtained by comparing the distributions of the $64$ codons in genomes, which agrees with the recruitment order of codons on the roadmap. This is a simple phylogenomic method to study the origins of metazoan, the evolution of primates, etc. This study should be regarded as an exploratory attempt to explain the diversification of the three domains of life by statistical mechanism in prebiotic sequence evolution. It is indicated that the number of bases in the triplet codons might be explained statistically by the number of strands in the triplex DNAs. The adaptation of life to the changing environment might be due to assembly of redundant genomes at the sequence level.}

\noindent {\bf Keywords:} the triplex2duplex picture $\vert$ genomic codon distribution $\vert$ universal genome format $\vert$ assembly of genomes $\vert$ the biodiversity space $\vert$ origin of the three-domains of life

\section{Introduction} 

Although there are always diverse of species, some common features of life such as the genetic code and homochirality remain invariant throughout the history of life. The unity of life for the contemporary species should rely on their common features at the molecular or sequence levels. A universal genome format at the sequence level is proposed in this article, which has been observed according to the common feature of genomic codon distributions for contemporary species. The formation of the universal genome format can be explained by a statistical model based on the assembly of genomes around the evolution of the genetic code. There are also distinguishing features of life in genomic codon distributions for diverse species besides the universal genome format in common. The main aim of this paper is to reconstruct the three-domain tree of life by comparing the distinguishing features of life in genomic codon distributions for diverse species and furthermore to explain the diversification of life. It is indicated that the separation of the three domains of life might be explained by groupings of codons based on the evolution of the genetic code. The statistical mechanism in the prebiotic sequence evolution is the key to understand the diversification of genomes in the main branches of the tree of life. 

There are some interesting relationships among the observations based on biochemical or genomic data at the molecular level or sequence level. The motivation of the three-part series of articles is to explain the remote yet profound relationship between the evolution of the genetic code and the tree of life. The theory at the sequence level in the present second part of this series of articles links the theory at the molecular level in the first part of this series (Li 2018-I) and the theory at the species level in the third part of this series (Li 2018-III). On one hand, the triplex picture for the prebiotic evolution was proposed in the first part of this series (Li 2018-I) to explain the concurrent origin of the genetic code. And the relationship between the evolution of the genetic code at the molecular level and the evolution of genomes at the sequence level need to be explained by the assembly of genomes, consisting of both coding and non-coding DNAs, for the three domains of life in the picture of transition from triplex nucleic acids to duplex nucleic acids (the triplex2duplex picture for short). On the other hand, the picture of genome evolution in the universal genome format will be proposed in the third part of this series (Li 2018-III) to explain the adaptation strategy of life to the changing earth's surface environment, where the common features of life, namely the genetic code, the homochirality and the universal genome format, play essential roles at the species level. 

There might be a transition from the prebiotic sequence evolution based on triplex DNAs to the sequence evolution based on the duplex DNAs. Such a triplex2duplex picture in the early evolution of life accounts for two significant observations. First, the number bases in triplet codons as ``three'' might be due to the common divisor ``three'' among the lengths of oligonucleotides generated in the triplex DNA evolution for assembly of proteins from motifs (Fig 1). Second, the observed three base fluctuations in the genomic codon distributions can be explained based on the simulations of assembly of genomes by triplet codons in the incomplete codon subset of the $64$ codons, where the metabolisms played crucial roles in forming the major branches of the tree of life (Fig 2). 

The genomic codon distribution is a useful method, which will be strictly defined based on complete genome sequences. The features of genomic codon distribution for a species can be divided into two parts: the distinguishing features for respective species and the common feature (namely the universal genome format) for all the species. The three domain tree of life can be reconstructed according to the Euclidean distances of species in the biodiversity space (Fig 3d) that is defined by comparing the distinguishing features of genomic codon distributions based on the genetic code evolution roadmap (Fig 1a in Li 2018-I). It is indicated that the three domains of life as well as virus originated at different stages during the evolution of the genetic code. The universal genome format can remain invariant for genome duplications during the genome size evolution. So the unity of life throughout the long history of life relies on this invariant common feature of life at the sequence level. 

\subsection{Problems}

According to the substitutions of triplex base pairs along the roadmap in the first part of this series (Li 2018-I), the prebiotic sequence evolution cannot be regarded as a merely random process. When the substitutions of the triplex base pairs abode by an order of stability from weak to strong, some specific sequences can be efficiently generated, while the other sequences can hardly be generated due to quite low efficiency. There were therefore some popular sequences that accumulated in the prebiotic surroundings, while other sequences that had not been generated can be regarded as the forbidden sequences. So the generated sequences may consist of a formal language. In the triplex picture, the functions of RNA and the functions of proteins coevolved. For instances, tRNA and aaRS coevolved, and rRNA and ribosomal proteins coevolved. A few popular sequences may possess certain biological functions, which were selected according to RNA-protein interactions; and the evolutionary information can be recorded in DNA sequences. A significant and inevitable technique arose: a genomic format is essential to record the genetic information in DNA sequences. A universal genome format is essential for genomes to record the genetic information. In fact, a universal genome format has been explicitly observed for all the species in the three domains of life (Woese et al. 1990) and virus, based on analysis of complete genome sequences. 

The origin of the three domains of life and the status of virus need to be clarified, where the mechanism for the separation of the three domains of life need to be explained at both the molecular and sequence levels. The relationship between the formation of biodiversity and the evolution of the genetic code need to be clarified. 

\subsection{Data and observations} 

The results in this article are mainly based on the complete genome sequences in GenBank. The animo acid frequencies of complete genome organisms are from GenBank. The classification of life is based on the NCBI taxonomy. The abbreviations for the taxa are as follows: Eukarya (E), Euryarchaeota (AE), Crenarchaeota (AC), Alphaproteobacteria (BPa), Betaproteobacteria (BPb), Gammaproteobacteria (BPc), Deltaproteobacteria (BPd), Epsilonproteobacteria (BPe), Firmicutes (BF), Actinobacteria (BAct), Cyanobacteria (BCya), Chlorobi (BChl), Spirochaetes (BSpi), Chlamydiae/Verrucomicrobia (BVer), Thermotogae (BThe), other bacteria (Bother), dsDNA virus (VdD), ssRNA virus (VsR), Mammalian orthoreovirus (VdRm), ssDNA virus (VsD), Blainvillea yellow spot virus (VdDB), satellite virus (Vsat). 

Genomic codon distribution is the main method for the statistical analysis of complete genome sequences. It is defined as a set of $64$ normalised distributions of codon interval distances in the complete genome sequence for a certain species. Based on observations of genomic codon distributions, several powerful notions are developed, such as the universal genome format, the distinguishing features in genome sequences for species, and the biodiversity space. 

Based on the genomic codon distributions, on the one hand, the tree of life can be obtained by averaging for codons and, on the other hand, the tree of codons can be obtained by averaging for species. It indicates the profound relationship between the phylogeny of contemporary species and the prebiotic evolution of the genetic code. 

As far as the common feature of genomic codon distributions is concerned, a universal genome format has been found for all the species, including viruses, bacteria, archaea and eukaryotes. As far as the distinguishing features of genomic codon distributions are concerned, some distinguishing features have been found for species, or for the three domains and virus, as well as for the lower taxa. These distinguishing features indicate the evolutionary relationships of species. 

\subsection{Main results} 

The universal genome format originated at the sequence level by taking advantage of statistical features in sequence evolution during the transition from the prebiotic sequence evolution in the triplex picture to the sequence evolution of duplex DNAs. The oligonucleotides can be generated in the prebiotic sequence evolution in accompany with the recruitment of codons in different incomplete subsets of the total $64$ codons. These oligonucleotides can be concatenated to form longer genes. Consequently, the primordial genomes can be assembled from these primordial genes. Hence the biological diversity formed and existed at the beginning of life due to the diverse assembly of genomes based on different incomplete subsets of codons at different stages of the evolution of the genetic code. The three domains of life as well as virus originated during different stages along the genetic code evolution roadmap.

The universal genome format is the common feature of genomic codon distributions for contemporary species. The three-base periodic fluctuations of the universal genome format can be explained by a sequence evolution model based on random assembly of codons from incomplete subset of the $64$ codons. The ``triplet'' in the genetic triplet code actually originated from the ``triplex'' in triplex DNA. A triplex DNA tends to form a single-stranded DNA whose length is three times the length of the three-stranded DNA. Considering the primordial translation mechanism, diverse protein motifs correspond to different triplex DNA sequences with different lengths. Then a gene for protein can be assembled by concatenating these oligonucleotides for motifs whose length is three times the length of the triplex DNA. In this process, the common divisor, namely ``three'', is chosen as the length of codons for the genetic triplet code, so as to avoid the frameshift mutations at junctions (Fig 1). The analysis of conserved protein structural features, such as polymer flexibility and loop closure, revealed that the earliest proteins consist of sequence modules whose typical size is $25 \sim 30$ amino acid residues (Berezovsky et al. 2000; Trifonov and Berezovsky 2003; Berezovsky et al. 2003; Aharonovsky 2005). This supports the assembly of proteins from short to long in the triplex2duplex picture (Fig 1). 

The biodiversification of the three domains and virus might be explained by the roadmap theory based on genomic data. Three notions, namely route bias, hierarchy bias and fluctuation amplitude, are introduced by comparing the genomic codon distributions according to the recruitment order of codons on the roadmap. Accordingly, a biodiversity space is defined by a three-dimensional Euclidean space with the three coordinates: route bias, hierarchy bias and fluctuation amplitude. A distance matrix is obtained by calculating the Euclidean distances between species in the biodiversity space, and consequently the tree of life can be reconstructed based on this distance matrix. This result supports the three domain tree (Woese 1990) rather than the eocyte tree (Lake et al. 1984), where the relationship between Euryarchaeota and Crenarchaeota is closer than the relationship between Crenarchaeota and Eukarytes. It is indicated that viruses were the earliest life on the earth. Virus differs from the three domains because their genomes belong to the type of genomes that can be assembled by earlier and smaller subsets of codons during the early evolution along the roadmap. It is also suggested that the diversification between bacteria and archaea is due to the leaf node bias according to the roadmap, and the diversification between archaea and eukaryota is due to the fluctuation amplitude. The separation between the unicellular and the multicellular organisms may originated during the assembly of eukaryotic genomes based on quasi-complete codon subset of the genetic code. This is a general phylogenomic method to obtain the tree of life based on biodiversity space. The result obtained based on the biodiversity space tends to support the three-stage pattern in Metazoan origins. 

\section{The picture at sequence level}

The prebiotic evolution may be explained in a hypothetical triplex picture in the first part of this series (Li 2018-I). There might be a transition from the triplex picture to the present duplex picture. During this triplex2duplex transition, some significant properties of life formed, such as the genetic triplet code and the universal genome format. The origin of molecules with certain biological functions need to be explained at both the molecular level and the sequence level. Though the homochiral system of life originated at the molecular level and the adaption of life formed at the species level, the sequence level bridges the gap between the molecular level and the species level. The origin of genomes is a milestone for the formation of living organisms out of the non-living ordinary matter during the evolution of the genetic code. Diverse genomes can be assembled by the non-coding and coding DNAs at different stages of the evolution of the genetic code along the roadmap. Thus, the evolution of the genetic code substantially determines the main branches of the tree of life. 

\subsection{Sequence evolution}

The codon degeneracy can be obtained according to the roadmap in the first part of this series (Li 2018-I) in a hypothetical triplex picture. The primordial sequence evolution of some small proteins and RNAs, such as the aaRSs and tRNAs, might be explained in the triplex picture. But the subsequent more complex and large ones need to be explained in the triplex2duplex picture. 

The substitutions of the triplex base pairs determine not only the evolution of the triplet codons but also the whole sequences of the triplex DNA. The difference in stabilities of the triplex base pairs accounts for the difference in the probability of substitutions of triplex base pairs. So the sequences were generated at different opportunities: some can be generated frequently, while others seldom. An RNA ribozyme or a primordial proteinase with certain biological function can be generated if its correspondence DNA was just among those frequently generated sequences. Not many primordial proteinases can be generated due to the low efficiency of the primordial translation mechanism. Once metabolic cycles were formed with the help of primordial proteinases, more proteinases can be produced in batch and can accumulated in the primordial surroundings. When the efficient modern translation mechanism was established, more and more proteinases can be produced in batches. The evolution of life can be boosted by the coevolution between RNA and proteins. The sequence evolution along the roadmap is by no means random. The substitution process can be reproduced along the roadmap, so the generation of certain sequences can repeat again and again. The prebiotic evolution was able to repeat along the roadmap again and again even if the prebiotic sequence evolution was interrupted by environmental harmful factors. 

There was a post-initiation-stage stagnation for the sequence evolution along the roadmap, during which the phase I amino acids involved to phase II amino acids (Wong 2005; Trifonov et al. 2006). Arbitrary finite sequences can be generated around the end of the initiation stage of the roadmap, which provided conditions for the generation of ribozymes and proteinases, such as tRNAs and aaRSs, and accordingly for the consequent expansion of the genetic code. The $GU\mbox{-}AG$ rule etc. (Breathnach et al. 1978; Padgett et al. 1986; Jackson 1991; Hall and Padgett 1994) might be explained by the post-initiation-stage stagnation phenomenon, where the self-splicing of introns played the role of topoisomerase (Fig 2). It is indicated that the earliest genes came from the purine $R$ strands on the roadmap considering that both the start and stop codons, even the non-standard stop codons, situated in the $R$ strand. The purine $R$ strands and pyrimidine $Y$ strands evolved relatively independently along the roadmap, which might relate to the origin of gender differences for the two-stranded DNA system. 

\subsection{From triplex DNA to duplex DNA} 

The stabilities of various sequences generated by the roadmap are different due to the differences in stability of triplex base pairs (Soyfer and Potaman 1996; Belotserkovskii et al. 1990). The single DNA strands, each of which consists of same base, tend to form a triplex DNA. However the single DNA strands, each of which consists of variety of bases, tend to form a duplex DNA. Starting from the $YR*R$ triplex DNA $Poly\ C \cdot Poly\ G * Poly\ G$, the sequence evolution of a triplex DNA had to stop up to a certain sequence, after which a duplex DNA tended to formed, instead of a triplex DNA. The neighbouring $5'$-end and $3'$-end bases in the anti-parallel strands can concatenate. Hence a triplex DNA can concatenate into a long single strand DNA whose length is three times the length of the triplex DNA. Furthermore, a duplex DNA can form by taking the above long single strand DNA as a template. Thus, a triplex DNA eventually evolved into a duplex DNA whose length had tripled. These long single-stranded DNA can encode proteins with certain biological functions, which can accumulate and form diversity of genomes. The universal genome format and the genetic triplet code can be explained in this triplex2duplex picture. 

I have done an experiment with small magnetic beads having the same sizes and same magnetism to simulate the transition from triplex DNA to duplex DNA, which stand for the polar molecules in the informative molecules. The polarity of the magnetic beads plays crucial role in the formation of a linear chain of magnetic beads. The strand direction in the triplex DNA can be simulated by the magnetic field direction in the chain of magnetic beads. There are hundreds of magnetic beads in a pile at the beginning. The generation of a single-stranded DNA can be easily simulated by pulling a chain of magnetic beads out of the pile of magnetic beads, where the polarity of the magnetic beads helps in forming the single-stranded chain of magnetic beads. The formation of three-stranded DNA by taking the single-stranded DNA as a template can be simulated by adding magnetic beads one by one in parallel or anti-parallel directions along the above single-stranded chain of magnetic beads. The anti-parallel directions of neighbouring strands of the $YR*R$ triplex DNA can be simulated by the anti-parallel magnetic field directions in the three-stranded magnetic bead chains as follows. Owing to the magnetic attraction between the neighbouring end beads of the anti-parallel strands, two magnetic beads at the end of the anti-parallel chains can easily concatenate to form a single-stranded chain of magnetic beads; and owing to the magnetic attraction between the other pair of neighbouring end beads, the third chain of magnetic beads can also join the single-stranded chain of magnetic beads. This experiment by magnetic beads indicates that it is easy to form a single-stranded chain from the three-stranded chain due to the concatenation at ends. Hence a long chain of magnetic beads has formed whose length has tripled. Furthermore, the formation of a double DNA can be simulated by adding magnetic beads one by one to the long chain of magnetic beads. Thus, the transition from the triplex DNA to duplex DNA has been simulated by the transition from three-stranded magnetic beads to two-stranded magnetic beads. The formation of the universal genome format and the genetic triplet code can also be explained by this simulation of the triplex2duplex picture. 

\section{Genomic codon distributions} 

The genomic codon distribution is an important method in this article, based on which the universal genome format, the distinguishing features of genomes for species and the biodiversity space are defined. 

\subsection{Definition} 

The genomic codon distribution for a species is a set of $64$ normalised distributions of codon intervals in the complete genome sequences, which is defined as follows. 

The $64$ codons $N'N''N'''$ ($N', N'', N''' = G, C, A, T$) correspond to $64$ distributions of codon intervals for a species. For each codon, a normalised distribution of codon intervals $dist(int, cod, spec)$ is defined based on the distribution of the interval distances of such a codon $cod$ in the complete genome sequence of a certain species $spec$. Concretely speaking, the base position $pos\_b=1, ..., gsize$ denotes the positions of bases from $5'$-end to $3'$-end in the complete genome sequence of the species $spec$ whose genome size is $gsize$. If a genome consists of several chromosomes, the complete genome sequence of this species consists of the sequences of all the chromosomes. For a certain species $spec$ and for a certain codon $cod=N'N''N'''$, the codon positions $pos\_c=pos\_c(1), pos\_c(2), ..., pos\_c(n\_c)$ denotes all the $n\_c$ positions of the codon $cod$ in the complete genome sequence of this species in the order from $5'$-end to $3'$-end, where $pos\_c(k)$ denotes that three bases at $pos\_c(k)$, $pos\_c(k)+1$ and $pos\_c(k)+2$ are respectively the same as the three bases of the codon $cod=N'N''N'''$. And the codon interval is define as $int\_c(k)=pos\_c(k+1)-pos\_c(k)$, where $k=1, ..., n\_c-1$. The codon interval distribution $dist\_int(int, cod, spec)=n\_int$ ($int=1, ..., cutoff$, it is accurate enough to choose $cutoff=1000$) for the codon $cod$ and the species $spec$ is defined by that there are totally $n\_int$ codon intervals among all the $n\_c-1$ codon intervals $int\_c(k)$ whose values are just equal to $int$. Furthermore, a normalised codon interval distribution is defined by $$dist(int, cod, spec)=\frac{dist\_int(int, cod, spec)}{\sqrt{\sum_{int} dist\_int^2(int, cod, spec)}}.$$ Thus, the genomic codon distribution for species $spec$ is defined by a set of the $64$ normalised distributions of codon intervals $dist(int, cod, spec)$. And the genomic codon distribution for a certain taxon is defined by the average of all the genomic codon distributions of different species in this taxon.
 
\subsection{Properties}

The codon interval distances for different codons are similar for a same species. An example is given to demonstrate the genomic codon distribution by taking the complete genome of Duck circovirus and the codons $GCA$ and $TAT$ for example (Fig 4c). First, all the positions of $GCA$ and $TAT$ in the genome are marked respectively; second, the interval distances for $GCA$ and $TAT$ are calculated respectively, and accordingly the distributions of intervals for $GCA$ and $TAT$ are obtained respectively, where the cutoff is set as $90$; and third, the genomic codon distribution for this virus is illustrated in two subfigures for the two codons $GCA$ and $TAT$ among the $64$ codons. It is observed that the normalised codon interval distances for different codons are similar for same species but are different for different species (Fig 5b). 

The genomic codon distribution obeys an additivity rule in genome duplications. So the features of genomic codon distribution is invariant in genome duplication, which played significant roles in genome evolution. Owing to the normalisation in the definition, the genomic codon distribution is independent of the genome size, which eliminates the influence of the immense difference in genome sizes among species (Gregory 2005, 2007; Plant DNA C-values database). In the present case, the distribution in the left subfigure and the distribution for genome duplication in the right subfigure are the same after normalisation (Fig 4d). In this sense, the genomic codon distribution is an invariant feature of a species, which ensures the invariance of the genomic codon distributions during the duplications or large-scale duplications in genome evolution. The universal genome format is a common feature of genomic codon distributions for all species. And there are also distinguishing features of genomic codon distributions for different species, by which the evolutionary relationship can be revealed. 

\section{Universal genome format}

\subsection{Origin of the triplet code}

In the first part of this series (Li 2018-I), an argument was given for the origin of the genetic triplet code according to the three-point fixation principle in the primordial translation mechanism. However, this significant problem need to be solved in a statistical manner in the triplex2duplex picture at the sequence level. 

The short DNA sequences generated in the triplex picture concatenated to long DNA sequences in the triplex2duplex picture. In this elongation process for both DNA sequences and protein sequences, the biological functions of short protein motifs acquired in the triplex picture should maintain during the elongation of protein sequences from short motifs to long proteins. Meanwhile, the fixation of the number of bases in codons as ``three'' is due to a statistical mechanism in translation from DNA to protein rather than any compulsory mechanism. Although there is no direct physical effect to set the number of bases in anti-codons of tRNAs as ``three'', the indirect statistical effect is valid to eventually choose the genetic triplet code, because it is the most effective solution for the continuous evolution of biological functions during the elongation of both DNA and protein sequences in the triplex2duplex picture. 

In the early triplex DNA evolution, the genetic triplet code had not formed yet. So the ``triplet'' code was not necessary for the synthesis of the earliest proteins, such as the early aaRSs, by the junior stage of the primordial translation mechanism in the first part of this series (Li 2018-I). But, in the late sequence evolution for both DNAs and proteins in the triplex picture, the triplex DNA tended to concatenate to single-stranded DNA whose length tripled. The common divisor of these tripled lengths is ``three''. In a statistical sense, the benefit of the choice of triplet codon is to maintain the prebiotic functions of motifs effectively (Fig 1). Taking the common divisor as the step length in reading mRNA in the translation mechanism is able to avoid the frameshift mutations in the assembly of proteins by motifs (Fig 1). Thus the biological functions of sequences evolved incessantly during the transition from the triplex DNA to the duplex DNA and furthermore during the elongation of sequences in the following evolution. So the choice of ``triplet'' code is due to the coevolution between DNAs and proteins, in a positive feedback manner, in the triplex2duplex picture. 

Both the genetic triplet code and the universal genome format can be explained in a statistical manner in the triplex2duplex picture. Actually, the formation of the universal genome format also facilitated the formation of the genetic triplet code, and vice versa. The three-base fluctuations of the universal genome format provides the respective average codon intervals for the $20$ cognate tRNAs to transport the $20$ canonical amino acids. This is an efficient recognition mechanism between tRNAs and an mRNA. The genetic triplet code between DNA sequences and protein sequences brought about the efficient universal genome format for modern translation mechanism, and the modern translation mechanism brought about more DNAs and proteins with advanced biological functions in return. Thus, the ``triplet'' code was strictly fixed by the universal genome format for genome organisation in the efficient modern translation mechanism. 

\subsection{Genome organisation}

According to the observation of genomic codon distributions based on the complete genome sequences of species in GenBank, there is a common three-base-fluctuation feature in all the genomic codon distributions of species, which is called as the universal genome format. There are fewer neighbouring codons with long codon intervals, so the peaks of the three base periodic fluctuations in the genomic codon distributions decline with the codon intervals. The universal genome format can be observed in the genomic codon distributions for species (Fig 5b). And it can also be observed in the genomic codon distributions for the three domains and virus and for the lower taxa (Fig 3b, 5a). Such an observed universal genome format can be explained, by statistical consideration, in the triplex2duplex picture based on the multi-staged evolution of the genetic code. 

There is a close relationship between the assembly of primordial genomes and the evolution of the genetic code. The observed universal genomic format can be explained based on the assembly of genomes during the multi-staged coevolution of the genetic code and the relevant molecules. The short DNA and protein sequences coevolved with the genetic code. Some short sequences evolved at the early stage of the genetic code evolution when only early codons had been recruited. Other short sequences evolved at the late stage of the genetic code evolution when additional codons had been recruited. Generally speaking, the early short sequences consisted of fewer early codons, while the late short sequences consisted of more codons. These short sequences generated at different stages in the evolution of the genetic code only consisted of codons in different incomplete subsets of codons rather than the complete set of all the $64$ codons. Furthermore, the short sequences can concatenate to long sequences in the triplex2duplex picture (Fig 1). So the genomes assembled by the early sequences consisted of fewer early codons, while the genomes assembled by the late sequences consisted of more codons. Thus there were diverse of primordial genomes, which corresponded to diverse of species at the beginning of life. Both the universal genome format and the distinguishing features of genomic codon distributions can be explained according to the assembly of genomes. Simulations have been developed based on the roadmap in the first part of this series (Li 2018-I) to explain the universal genome format and the distinguishing feathers of genomic codon distributions. The simulations agrees well with the observations. 

The universal genome format can be explained by the transition from triplex DNA to duplex DNA. The short DNA segments consist of codons that are chosen randomly from an incomplete subset of codons, and the long DNA sequences were concatenated by the short DNA segments, and then the genomes were assembled by the long DNA sequences in the triplex2duplex picture. The origin of the triplet codon depends on the invariance of the functions of motifs, which also resulted in the three base periodic fluctuations in genomic codon distributions. The additivity of the universal genome format ensures the invariance of genome features in genome duplications. The universal genome format is also a local property, whose features also apply to parts of the genome of a species. The transposition process can be taken as the duplication of a local part of a genome, so the genomic codon distribution is invariant in the popular transposition processes. The universal genome format also remains unchanged in horizontal gene transfer due to the universality of this format among species. 

When studying the distributions of other kinds of segment intervals based on genomic data in GenBank, for example, the distribution of $2$-base segment intervals, the distribution of $4$-base segment intervals or the distribution of $5$-base segment intervals, other than the distribution of ``triplet'' codon intervals defined in subsection 3.1, the periods of the fluctuations in the above distributions of other kinds of segment intervals are always ``three'', rather than $2$, $4$ or $5$. So the specific number ``three'' in the three base periodic fluctuations of the universal genome format indeed relates to the number ``three'' in the genetic ``triplet'' codons. The universal genome format is a conservative property in the sequence evolution, which has been maintained from generation to generation. The universal genome format applies to the three domains and viruses. This indicates that the universal genome format appeared before the separation of the three domain and virus. For the same reason, the origin of the genetic code appeared before the separation of the three domain and virus, too. Therefore, in the formation of primordial genomes, the features in sequences were established in terms of the common features of life, such as the three base periodic fluctuations in the universal genome format and the genetic triplet codons. 

\subsection{Simulation of universal format}

The universal genome format can be simulated by a statistical model for genome assembly in the triplex2duplex picture, which includes the simulation of the concatenation of the DNA sequences based on incomplete subset of codons and the simulation of the genome assembly. The simulation procedure are described as follows. 

An incomplete subset of codons is chosen from the $64$ codons, whose number cannot be too small, or too large to be near to $64$. Some codons are chosen randomly one by one from this incomplete subset. A short DNA sequence can be generated by concatenating these codons. Hence many short DNA sequences can be generated for a certain incomplete subset of codons. Furthermore, other short DNA sequences can also be generated for other incomplete subset of codons. And the long DNA sequences can be obtained by concatenating these short DNA sequences. A simulated genome can be assembled by concatenating all these long DNA sequences. Hence, the simulated genomic codon distribution is obtained by calculation based on the simulated genome sequences. The universal genome format, namely the three base periodic fluctuations, has been observed in the simulated genomic codon distributions. The simulated genome sequence is not random if choosing codons from incomplete subset of codons. However, if choosing codons from all the complete set of all the $64$ codons, the three base periodic fluctuations will vanish; namely the simulated genomic codon distributions become smoothly declining curves, which becomes a random distribution. 

\section{Statistical features of genomes}
\subsection{Virus, bacteria, archaea, eukarya}

The features of genomic codon distributions are different among Virus, Bacteria, Archaea and Eukarya (Fig 3b,
5b). The genomic codon distribution for a domain is obtained by averaging the genomic codon distributions of all the species in this domain (only considering those with complete genome sequences, in practice) (Fig 3b). The feature of genomic codon distribution for a species is similar to that of the corresponding domain (Fig 3b, 5b). The method based on the genomic codon distribution apply not only for cellular life but also for viruses. Explanation of the features of the genomic codon distributions (Fig 3b, 5c) is helpful to understand the origin of the three domains and to clarify the status of virus.

The features of genomic codon distributions for the domains can be described respectively from two aspects: first, the periodicity and the amplitudes of the genomic codon distributions; second, the order of the average distribution heights for $64$ codons. ($i$) The features of genomic codon distribution for Virus include: declining three-base-periodic fluctuations, high amplitudes (Fig 3b, 5b). In addition, the order of the average distribution heights is: $Hierarchy\ 1$, highest; followed by $Hierarchy\ 2$ and $Hierarchy\ 3$; and $Hierarchy\ 4$, lowest (Fig 6a). ($ii$) The features of genomic codon distribution for Bacteria include: declining three-base-periodic fluctuations, medium amplitudes (Fig 3b, 5b). In addition, the order of the average distribution heights is: $Hierarchy\ 1$, highest; followed by $Hierarchy\ 2$, $Hierarchy\ 3$ and $Hierarchy\ 4$ (Fig 6a). ($iii$) The features of genomic codon distribution for Archaea include: declining three-base-periodic fluctuations, medium amplitudes (Fig 3b, 5b). In addition, the order of the average distribution heights is roughly opposite to that of Bacteria: $Hierarchy\ 4$, highest; followed by $Hierarchy\ 3$ and $Hierarchy\ 2$; and $Hierarchy\ 1$, lowest (Fig 6a). Comparing Archaea with Bacteria, the dwarf 2nd peak at 6-base interval in the genomic codon distribution is even lower than the 3rd peak at 9-base interval (Fig 3b, 5b). This detailed feature is different from Bacteria. For Bacteria, the 2nd peak is higher than the 3rd peak for codons in $Hierarchy\ 1$ and the heights are almost the same for the 2nd and 3rd peaks for other codons (Fig 3b, 5b). ($iv$) The features of genomic codon distribution for Eukarya are more complicated than for other domains, which include: declining distributions, indistinct multi-base-periodic fluctuations, ups and downs of peaks, or overlapping peaks for some codons (Fig 3b, 5b). In addition, the order of the average distribution heights is similar to that of Archaea: $Hierarchy\ 4$, highest; followed by $Hierarchy\ 3$ and $Hierarchy\ 2$; and $Hierarchy\ 1$, lowest (Fig 6a). The features of genomic codon distribution for Archaea put together the features of both Archaea and Eukarya (Fig 3b, 5b). Crenarchaeota and Euryarchaeota are similar in genome size distributions (Fig 3b, 5a). It is indicated that the three domain tree is more reasonable than the eocyte tree of life, from the viewpoint on the features of genomic codon distributions (Fig 3b, 5a).

The origin order for the domains can be obtained according to the linear regression plot whose abscissa and ordinate represent the recruitment order of codon pairs and the order of average distribution heights respectively (Fig 3c, 6a). The greater the slope of the regression line for a domain is, the more the late recruited codons are. So, the order of the slopes of the regression lines represents the origin order of the domains: $1$ Virus, $2$ Bacteria, $3$ Archaea, $4$ Eukarya (Fig 3c). The slope for Virus is negative and least (Fig 3c), which means that majority of codons in the sequences recruited early. The virus originated earlier than the three domains. The slope for Bacteria is negative (Fig 3c), and the proportion of codons in $Hierarchy\ 1$ for Bacteria is high; the proportion of codons in $Hierarchy\ 2 \sim 4$ for Bacteria is lower (Fig 6a). The slope for Archaea is positive (Fig 3c), which means that majority of codons in the sequences recruited late. The slope for Eukarya is positive and greatest (Fig 3c), which also means that majority of codons in the sequences recruited late. The Eukarya originated latest. Archaea originated between Bacteria and Eukarya (Fig 3c). Limited data indicate that the slope for Crenarchaeota is less than the slope for Euryarchaeota (Fig 3c), which suggests that Crenarchaeota may originate earlier than Euryarchaeota. This origin order of the domains is independent from the base compositions for the domains, which are different from the order of the domains by $GC$ content (Fig 4b). Note that some Eukaryotic genomes mainly consists of non-coding DNAs, the tree of eukaryotes based on their genomic codon distributions is mainly depended on the non-coding DNA (Fig 4f, 7j). This indicates that the non-coding DNA has played a substantial role in the evolution of eukaryotes. 

According to the roadmap, the initial subset took shape in the initiation stage of the evolution of the genetic code (Fig 1a in Li 2018-I), where the $GC$ content of the initial subset was $13/18$; leaf nodes appeared in the ending stage (Fig 1a in Li 2018-I), where the $GC$ content of the leaf nodes decreased to $16/48$ (Fig 4a). The variation range of $GC$ contents from $16/48$ to $13/18$ obtained by the roadmap generally agrees with the variation range of $GC$ contents in observation for contemporary species (Fig 4a, 4b). The $GC$ content can vary in a common wide range for taxa such as virus, bacteria, archaea, eukarya. So the variation of $GC$ content is the consequence rather than the cause of biodiversification of the main branches of tree of life. 

\subsection{Simulation of distinguishing features}

The features of the genomic codon distributions for the domains can be simulated by a stochastic model. The features of declining three-base-periodic fluctuations (Fig 3b) can be simulated by the model (Fig 5c). In the model, the bases $G$, $C$, $A$, $T$ are given in certain base compositions, from which three bases are chosen statistically to make up a codon. Many a codon concatenates together to form a genome sequence in simulation. Hence, the genomic codon distribution is obtained by calculating the simulated genome sequence. When the length of the triplex DNA increased in the evolution along the roadmap (Fig 4a), more and more codons are accommodated into the triplex DNA. The feature of declining three-base-periodic fluctuations, therefore, is achieved in the simulations. Let $n_0$ denote the number of codons in the incomplete subset of codons. The $n_0$ codons are generated statistically to concatenate a simulated genome by several cycles. The genomic codon distribution thus obtained has the feature of declining three-base-periodic fluctuations. When $n_0$ is small (e.g., around $16$), the amplitude of the three-base-periodic fluctuations is great. The greater $n_0$ is, the less the amplitude of the three-base-periodic fluctuations is. When $n_0$ is $32$ (namely all codons participated in the concatenation), the three-base-periodic fluctuations vanish. The amplitude of the three-base-periodic fluctuations decreases from Virus to Bacteria, then Archaea, finally Eukarya (Fig 3b, 5c), which results from the increasing number of codons accommodated in the triplex DNA at different stages of the evolution of the genetic code. The complicated features of genomic codon distributions for eukaryotes can also be simulated by the model. The features of ups and downs of peaks for eukaryotes can be simulated by concatenating different set of codons that are generated from different subsets of the genetic code (Fig 5c). The simulated alternative ``high peak - low peak'' feature for some codons and the low amplitudes of the genomic codon distributions for other codons agree with the observations for true eukaryotes (Fig 3b, 5b, 5c).

The domains differ in the orders of the average distribution heights, which can also be simulated by the statistical model (Fig 6b). The $GC$ content tends to decrease along the evolutionary direction according to the roadmap (Fig 1a, 9b in Li 2018-I), which accounts for the above different orders among the domains. Different base compositions for $G$, $C$, $A$, $T$ are assigned for Virus, Bacteria, Archaea and Eukarya in the simulations, in consideration of the decrement of $GC$ content effected by the origin order of the domains. Different orders of the average distribution heights have been obtained (Fig 6b) based on the simulations of genomic codon distributions for the domains (Fig 5c). In the simulation, high $GC$ contents are assigned to Virus or Bacteria; hence the average distribution heights from high to low are as follows: $Hierarchy\ 1$, $Hierarchy\ 2$, $Hierarchy\ 3$, and $Hierarchy\ 4$ (Fig 6b). And low $GC$ contents are assigned to Archaea or Eukarya in the simulation; hence the average distribution heights from high to low are as follows: $Hierarchy\ 4$, $Hierarchy\ 3$, $Hierarchy\ 2$, and $Hierarchy\ 1$ (Fig 6b). The simulations generally agree with the observations as to the order of the average distribution heights (Fig 6a, 6b).

Incidentally, the validity of the composition vector tree (a method to infer the phylogeny from genome sequences) in the literatures (Deschavanne et al. 1999; Qi et al. 2004; Delsuc et al. 2005) can be explained based on the genomic codon distribution in this paper (Fig 5c). The average distribution heights for the $64$ codons are proportional to the numbers of codons in the genomes. The summations of the elements of the unnormalised genomic codon distributions for a species represent the numbers of the $64$ codons in the genome. So a set of average distribution heights is equivalent to the $K=3$ composition vector for this species. For example, the summations of the elements of the unnormalised genomic codon distributions of $GCA$ and $GCA$ are $32$ and $38$ respectively, which means that there are $33$ $GCA$ and $39$ $TAT$ in the genome (Fig 5c). Hence, the $K=3$ composition vector for this species is obtained, where the elements corresponding to $GCA$ and $TAT$ are $33$ and $39$ respectively (Fig 5c). The $K=3$ composition vector (Fig 5c) in fact corresponds to the order of the average distribution heights (Fig 6a). It has been emphasised that the genomic codon distributions are essential features of species at the sequence level. The composition vector, as a deformed method based on the genomic codon distribution, is certainly valid to infer the phylogeny.

\section{Biodiversity from the genetic code}

In the first part of this series (Li 2018-I), the roadmap theory was proposed, as a hypothesis, to explain the evolution of the genetic code and the origin of the homochirality of life. In this article, more experimental evidences are provided to enhance and reinforce the roadmap theory (Fig 3). Especially, the four hierarchies on the roadmap are directly distinguished in a tree of codons obtained by studying the genomic codon distributions (Fig 3a). And the origin of the three domains of life might be explained based on the roadmap and the genomic data (Fig 3d). In short, the biodiversity originated in the evolution of the genetic code, and has flourished by the environmental changes.

\subsection{Profound relationship}

The viewpoint that the tree of life was rooted in the recruitment of codons can be proved by analysing the complete genome sequences. The genomic codon distribution is a powerful method to study the statistical properties of the complete genome sequences, which reveals the profound relationship between the primordial genetic code evolution and the contemporary tree of life (Fig 3a, 4e). The genomic codon distribution for a species with complete genome sequence is defined as follows (see section 3.1 for detail): first, mark all the positions for each of the $64$ codons in its complete genome sequence respectively; second, calculate all the codon intervals between the adjacent marked positions for each of the $64$ codons respectively, and obtain the distributions of codon intervals according to the $64$ sets of codon intervals respectively; third, obtain the genomic codon distribution by normalising the $64$ distributions of codon intervals, which eliminates the direct impact of a wide range variation of genome sizes.

Both the tree of codons (Fig 3a) and the tree of species (Fig 4e) are obtained according to the same set of correlation coefficients among the genomic codon distributions. $952$ complete genomes in GeneBank (up to 2009) were used in the calculations. There are $64$ genomic codon distributions for each of the $952$ species. On one hand, the average correlation coefficient matrix for codons is obtained by averaging these $64 \times 64$ correlation coefficient matrices for the $952$ species; hence the tree of codons is obtained based on the $64 \times 64$ distance matrix that equals to half of one minus the average correlation coefficient matrix for codons (Fig 3a); on the other hand, the average correlation coefficient matrix for species is obtained by averaging these $952 \times 952$ correlation coefficient matrices for the $64$ codons, hence the tree of species is obtained based on the $952 \times 952$ distance matrix that equals to half of one minus the average correlation coefficient matrix for species (Fig 4e). The codons from the four hierarchies $Hierarchy\ 1 \sim 4$ on the roadmap gather together in the tree of codons respectively, where the $Hierarchy\ 1$ and $Hierarchy\ 2$ situated in one branch, and $Hierarchy\ 3$ and $Hierarchy\ 4$ in another branch (Fig 3a). This is a direct evidence to support the roadmap (Fig 1a in Li 2018-I). According to the NCBI taxonomy, the tree of species roughly presents the phylogeny of species (Fig 4e). Especially, the tree of the eukaryotes with complete genomes is reasonable to some extent (Fig 4f). In section 8.2, an improved tree of species (Fig 7h, 7j) will be obtained by dimension reduction of the $64 \times 952$ genomic codon distributions according to the roadmap for the evolution of the genetic code in the first part of this series (Li 2018-I) (Fig 5b, 3d).

\subsection{Feasibility}

It is also a remote relationship between the genetic code evolution about four billion years ago and the tree of life for contemporary species. The feasibility of the method to study the biodiversification by the evolution of the genetic code based on genomic codon distributions can be explained as follows. Starting from the triplex DNA $Poly\ C \cdot Poly\ G*Poly\ G$ in the triplex picture, base substitutions along the roadmap may occur at all the places throughout the whole triplex DNA (Fig 4a). The distinguishing features in the primordial sequence evolution formed and evolved with the evolution of the genetic code. More distinguishing features due to the transition from triplex DNA to duplex DNA also coevolved with the genetic code. Both the universal genome format and the distinguishing features of the genomic codon distributions were conservative due to additivity of genomic codon distributions during large scale duplications in genome evolution (Fig 4d). The whole history for assembly of genomes has been recorded in the genomic codon distributions of ancestors of species in respective taxa. Especially, the evolution of the genetic code has also been recorded in the genomic codon distributions. In the long-lasting history of the evolution of life, even if genome sizes have been doubled for many times, the distinguishing sequence features of the ancestors can be inherited generation by generation from the primordial period until present. The biodiversity at the species level essentially originated in the assembly of genomes during the evolution of the genetic code at the sequence level. The main branches of the tree of life closely relates to the features of genomic codon distributions formed during the evolution of the genetic code. Redundant genomes have been produced at the sequence level, the extinction and origination in the history of life do not change distinguishing sequence features of the ancestors, which just reduce or add the number of species. The main branches of the tree of life has always been preserved. Hence, it is not surprising that the ancient information on the evolution of the genetic code and the origin of the three domains can be dug out from the genomic codon distributions based on the complete genome sequences of contemporary species. 

\section{On the tree of life}
\subsection{The biodiversity space}

According to the roadmap theory and the features of genomic codon distributions, an biodiversity space (Fig 3d) has been constructed to explain the major and minor branches of the tree of life (Fig 7h, 7i). The biodiversity space is a 3-dimensional Euclidean space, where each species situates at a certain point. The three coordinates of the biodiversity space are $fluctuation$, $route\ bias$ and $hierarchy\ bias$ (Fig 3d), which are defined respectively as follows: ($i$) The fluctuation for a species is defined as the logarithm of the total summation of the absolute value of the neighbouring differences of the fluctuations in the genomic codon distributions for all the $64$ codons: 
$$fluctuation (spec) = \ln (\sum_{int=1 \cdots 999,\ cod=1\cdots 64} \Vert dist(int+1, cod, spec)-dist(int, cod, spec) \Vert );$$
($ii$) The route bias for a species is defined by the difference of the average codon weights of genomic codon distribution between $Route\ 0$ and $Route\ 1 \sim 3$ (Fig 7b): 
$$route\ bias (spec) =\frac{1}{48} \sum_{cod \in Route 1,2,3} cwdist(cod, spec) -  \frac{1}{16} \sum_{cod \in Route 0} cwdist(cod, spec);$$
($iii$) The hierarchy bias for a species is defined by the difference of the average codon weights of genomic codon distribution between $Hierarchy\ 1 \sim 2$ and $Hierarchy\ 3 \sim 4$ (Fig 7b): 
$$hierarchy\ bias (spec) = \frac{1}{32} \sum_{cod \in Hierarchy 3,4} cwdist(cod, spec) -  \frac{1}{32} \sum_{cod \in Hierarchy 1,2} cwdist(cod, spec);$$
where the codon weight of genomic codon distribution $cwdist(cod, spec)$ for codon $cod$ and species $spec$ is defined based on the genomic codon distributions as follows:
$$cwdist(cod, spec)= \sum_{int=1 \cdots 1000} dist^2(int, cod, spec) /  \sum_{int=1 \cdots 1000,\ cod=1\cdots 64} dist^2(int, cod, spec) .$$

It should be emphasised that the species in the three domains generally cannot be divided into well-separated clusters based on otherwise defined biases of average codon weights of genomic codon distribution between two arbitrarily divided groups of codons, such as between $cod \in odd$ and $cod \in even$, between $cod \in Y\ strands$ and $cod \in R\ strands$, and majority cases between $cod \in route\ i$ and $cod \in route\ j$ ($i,j=0 \cdots 3$), etc. So, the definitions of route bias and hierarchy bias themselves reveal the essential features of the evolution of the genetic code along the roadmap. After groping around, it was realised that only rare proper groupings of codons based on the roadmap in the first part of this series (Li 2018-I) are able to divide the species in the three domains (Fig 3d). Actually, the choice of route bias and hierarchy bias for the coordinates of the biodiversity space depends on the results to divide the species in the three domains into well-separated clusters, where the recruitment order of codons along the roadmap has been considered (Fig 7a, 7b). The genome size distributions can be divided into $20$ groups according to the $20$ combinations of the $4$ bases (Fig 7a). The number $20$ of combinations is equal to the number of canonical amino acids, which can improve the efficiency of recognition between tRNAs and mRNAs (6b in Li 2018-I). So, the classifications of genome size distributions are of biological significance. The definitions of route bias and hierarchy bias as well as fluctuation based on the recruitment of codons on the roadmap are motivated by biological considerations rather than mathematical expedience. 

The three coordinates of the biodiversity space are obtained by calculations for each species, thus a species corresponds to a point in the biodiversity space (Fig 3d). The species gather together in different clusters for Virus, Bacteria, Archaea and Eukarya respectively (Fig 3d, 7c, 7d, 7e). The cluster distributions of taxa of lower rank are also observed in detail in the biodiversity space (Fig 3d, 7c, 7d, 7e). The evolutionary relationships among species are thereby illustrated in the biodiversity space, which reveals the close relation between the origin of biodiversity and the origin of the genetic code. According to the fluctuations of genomic codon distributions, the order of domains is as follows: first Virus, then Bacteria and Archaea, and last Eukarya (Fig 7d, 7e). Bacteria and Archaea are distinguished roughly by route bias (Fig 7c, 7e), and they can be distinguished better by both hierarchy bias and route bias (Fig 3d, 7c). So the separation of the three domains is due to route bias and hierarchy bias at the sequence level. Furthermore, the leaf node bias can be defined by the difference of the average codon weights of genomic codon distributions between the leaf nodes in $Route\ 0 \sim 1$ and the leaf nodes in $Route\ 2 \sim 3$ (Fig 1a in Li 2018-I, 7b). Hence, Crenarchaeota and Euryarchaeota are distinguished by leaf node bias (Fig 7f, 7g). Diversification of Crenarchaeota and Euryarchaeota is, therefore, due to leaf node bias during the ending stage in the evolution of the genetic code. The cluster distributions of species from the three domains in the biodiversity space shows that the formation of biodiversity is due to the evolution of the genetic code at the sequence level. And the earlier stage of the evolution of the genetic code contribute more to the main branches of the tree of life.

\subsection{From tree of codons to tree of life}

A tree of life (Fig 7h) can be obtained based on the Euclidean distances among species in the biodiversity space (Fig 3d). The three coordinates of the biodiversity space need to be equally weighted in advance by non-dimensionalisation: $(X(spec)-\mu)/\sigma$, where $X(spec)$ is any of the three coordinates, $\mu$ is the mean value of $X(spec)$ and $\sigma$ is the standard deviation of $X(spec)$. Then a distance matrix is obtained by calculating the Euclidean distances among species in the non-dimensionalised biodiversity space. Thus, the tree of life is constructed by PHYLIP software based on the distance matrix (Fig 7h). Note that all the phylogenetic trees in this paper are constructed by PHYLIP software using UPGMA method (Felsenstein 1981). The tree of life based on the biodiversity space is reasonable to describe the evolutionary relationships among species. Take the eukaryotes for example. Such a tree provides reasonable inclusion relationships among vertebrates, mammals, primates, where chimpanzee is nearest to human beings. And yeast indicates its root (Fig 7j).

A tree of taxa can also be obtained according to the average distances of species among Virus, Bacteria, Crenarchaeota, Euryarchaeota and Eukarya in the non-dimensionalised biodiversity space (Fig 7i). The tree of taxa shows that the relationship between Crenarchaeota and Euryarchaeota is nearer than the relationship between Euryarchaeota and Eukarya (Fig 7i). The tree of taxa based on the biodiversity space agrees with the three domain tree rather than the eocyte tree (Woese 1990; Lake et al. 1984; Lake 1988; Archibald 2008; Cox et al. 2008). This result is accordant with the observations that features of genomic codon distributions between Crenarchaeota and Euryarchaeota (Fig 5a) are more similar than the features between Eukarya and Euryarchaeota (Fig 3b Eukaryote vs. Fig 5a).

There is a profound relationship between the evolution of the genetic code and the formation of biodiversity. The evolutionary relationship among codons on the roadmap plays significant roles in assembly of primordial genomes and accordingly in the separation of the three domains as well as virus. So both the tree of codons and the three domain tree of life can be obtained based on the conservative features of genomic codon distributions. 

\subsection{Origins of metazoan and evolution of primates}

The distances among species in the biodiversity space are obtained based on genomic codon distributions. The more similar in genomic codon distributions, the shorter distances in the biodiversity space, hence the closer evolutionary relationships (Fig 3d, 7c~7e). The tree of primates (Fig 8h) can be obtained based on comparing their genomic codon distributions (Fig 8a~8g, Fig 5b). Concretely, this tree is obtained as a consensus tree of the trees of primates based on respective homologous chromosomes. 

There is a disputation on the origins of metazoan, and a three-stage hypothesis on the origins of metazoan has been proposed based on fossil observations (Conway-Morris 1993, 1989; Budd and Jensen 2000; Valentine 2001; Shu 2005, 2008; Shu et al. 2004; Shu et al. 2009). Such a disputation on the origins of metazoan need to be evaluated from outside palaeontology. The evolutionary information for species are stored in their genomic codon distributions (Fig 9). The species in certain taxon roughly aligned linearly in the biodiversity space, which indicates some evolutionary information. It can be observed that the diploblastica, protostomia and deuterostomia cluster separately in the biodiversity space (Fig 10). This observation tends to support the three-stage pattern in Metazoan origins. 

\section{Conclusion and discussion}

The three domain tree of life has been reconstructed by comparing the distinguishing features in the genomic codon distributions based on groupings of codons on the roadmap for the evolution of genetic code, which indicates that there should be a close relationship between the diversification of life and the evolution of the genetic code. The common features as well as the distinguishing features in the genomic codon distributions for contemporary species have been simulated by a statistical model on prebiotic sequence evolution. This study is an exploratory attempt to explain the diversification of the three domains of life based on the evolution of the genetic code and the assembly of primordial genomes, in a statistical manner. The observations are based substantially on the biological data. It should be emphasised that this article is a hypothesis, and we need to take a critical look at the interpretations in this study.

\section*{Acknowledgements} My warm thanks to Jinyi Li for valuable discussions. I wish to thank the contributors of the biological data used in this study. Supported by the Fundamental Research Funds for the Central Universities.

\clearpage \begin{figure}
 \centering
 \includegraphics[width=18.3cm]{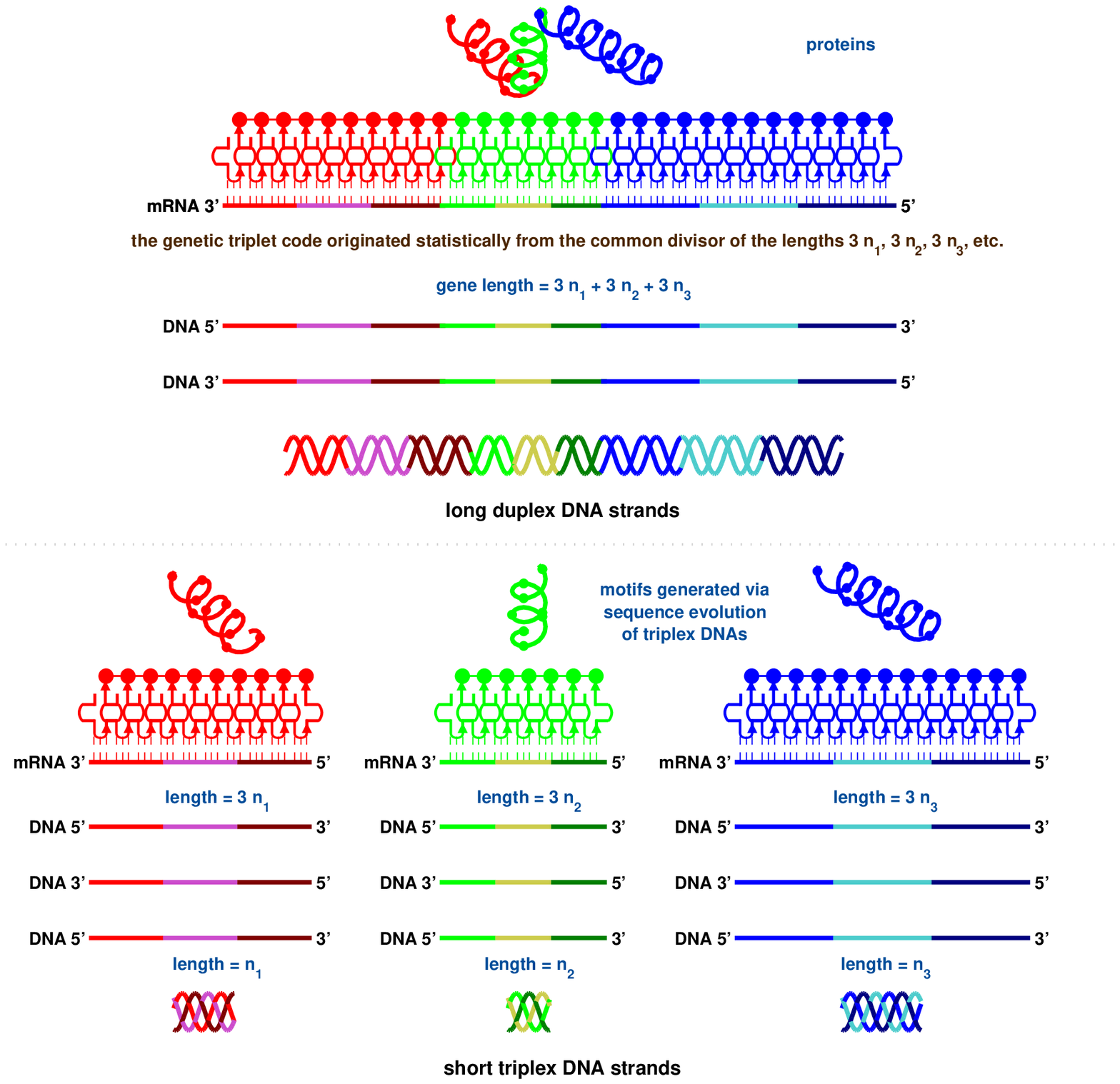}
 \caption{Explanation of the number of bases in codons as $3$ in the triplex2duplex picture. The primordial motifs were generated according to the short sequence evolution in the triplex picture. And the origin of protein is due to the assembly of gene in the duplex DNA from the motifs generated in the triplex DNA. The origin of the triplet codons is due to statistical dynamics in the assembly of genes. The functions of short motifs maintain in the long proteins unless there are breaks or overlaps in the seam areas between short sequences. }
\end{figure}

\clearpage \begin{figure}
 \centering
 \includegraphics[width=18.3cm]{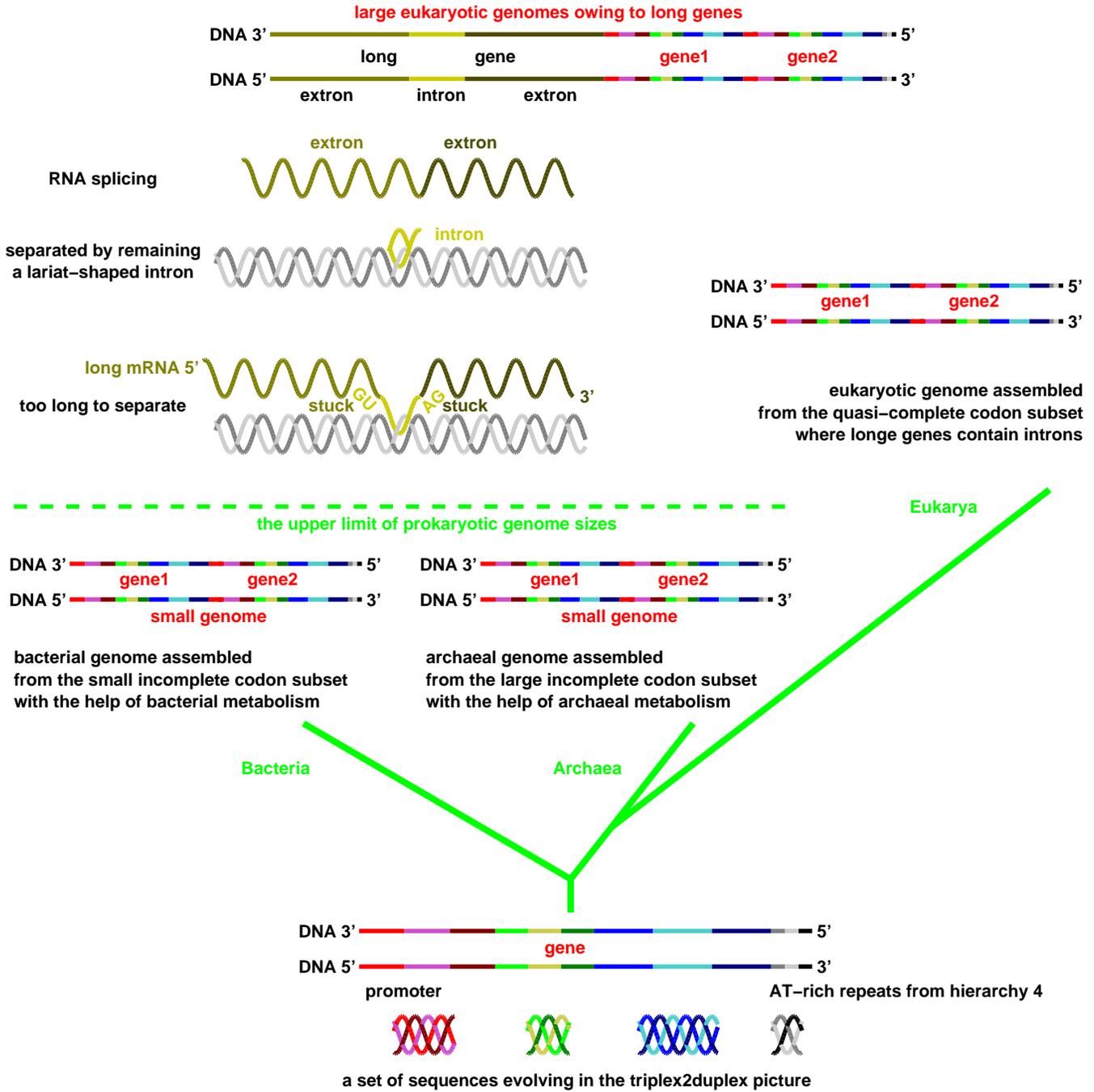}
 \caption{Explanation of the origin of the three domains according to the assembly of genomes. The universal genome format of the three-base periodic fluctuations is due to the assembly of genomes from the codons in incomplete codon subsets. The origin of long genes for eukaryotes is due to the topological function of lariat-shaped introns. }
\end{figure}

\clearpage \begin{figure}
  \centering
  {\small \bf a} \includegraphics[width=16cm]{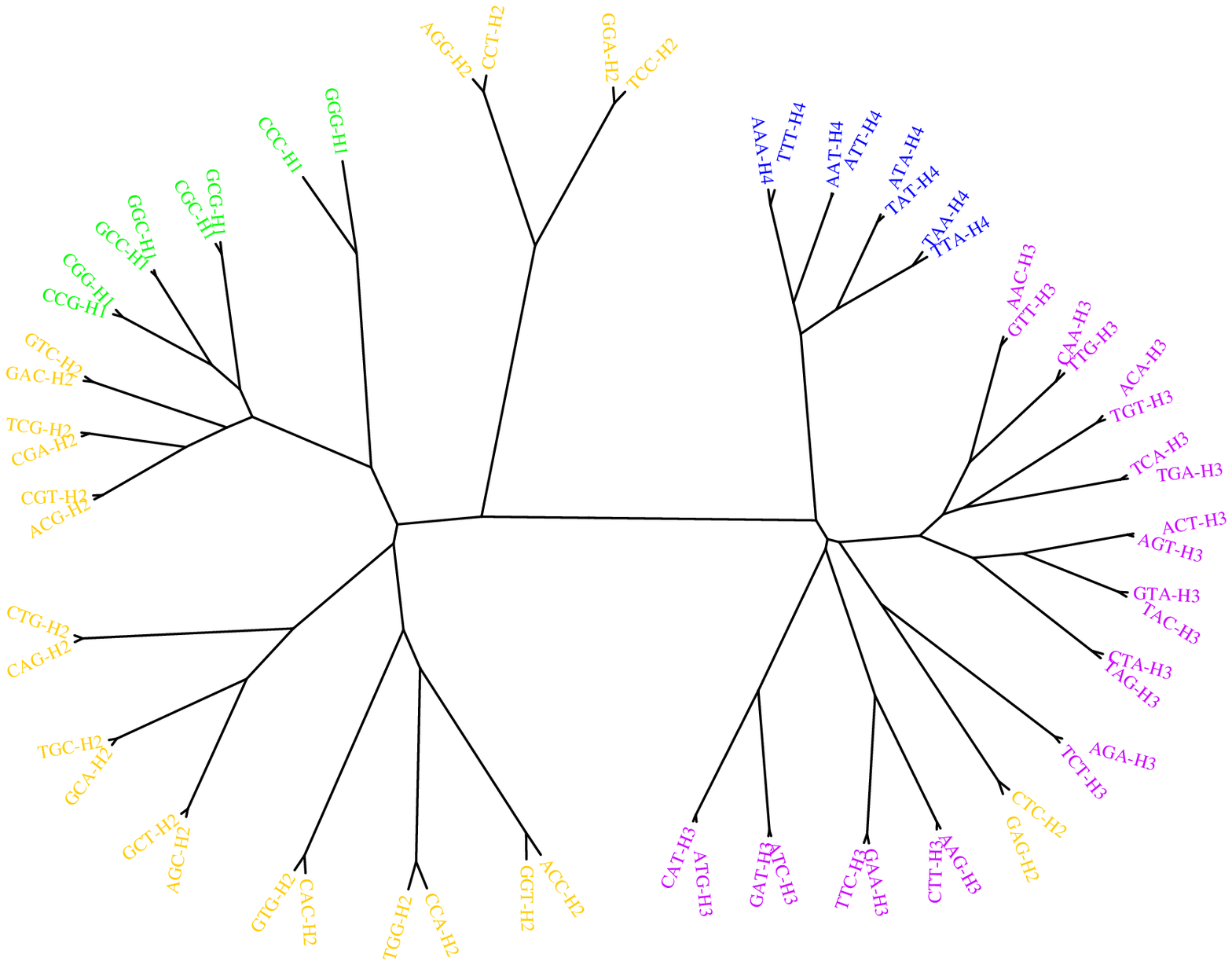}\\ \vskip 0.5cm
  {\small \bf b} \includegraphics[width=16.5cm]{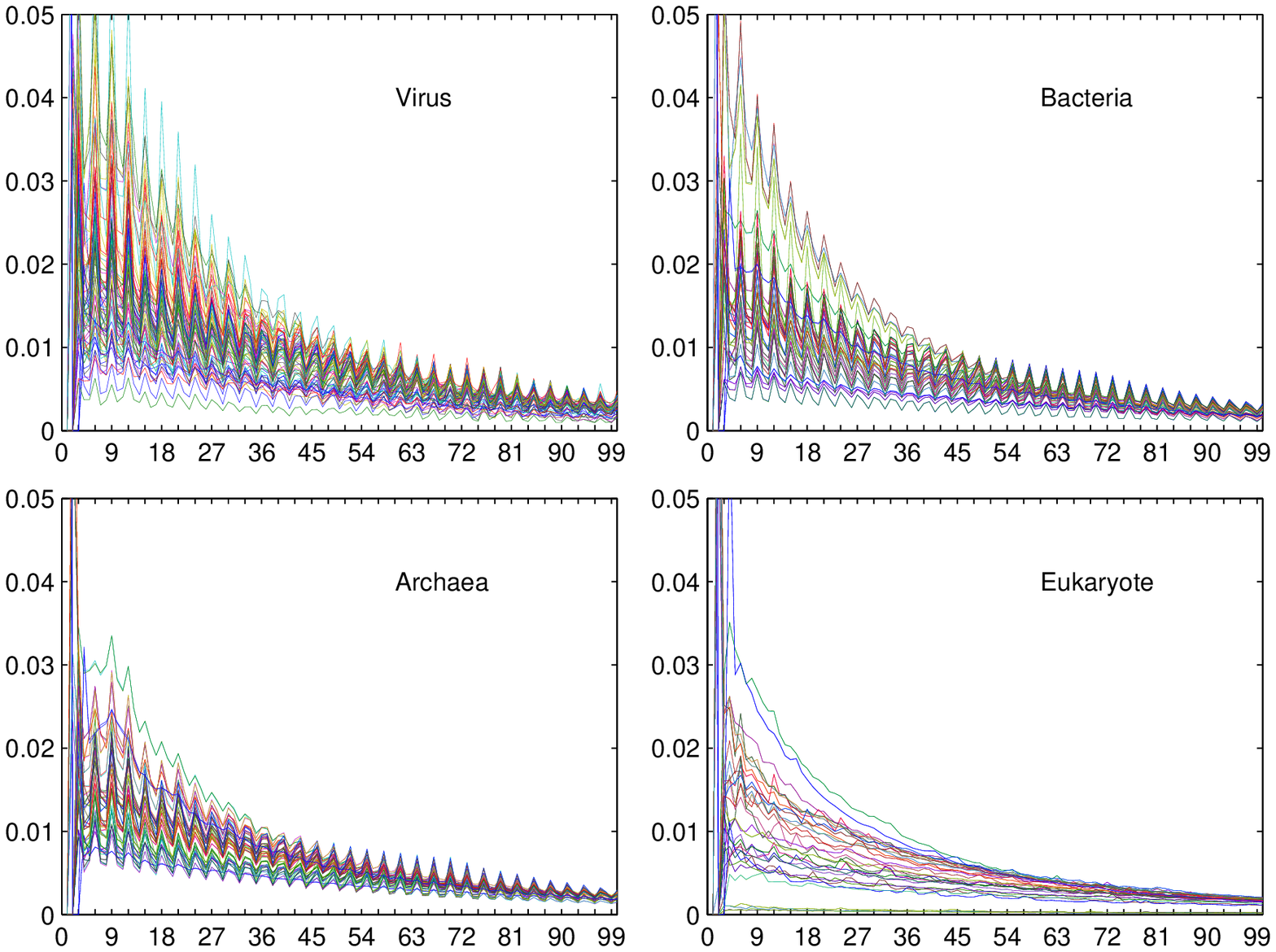}
\end{figure}

\clearpage \begin{figure}
  \centering
  {\small \bf c} \includegraphics[width=16.5cm]{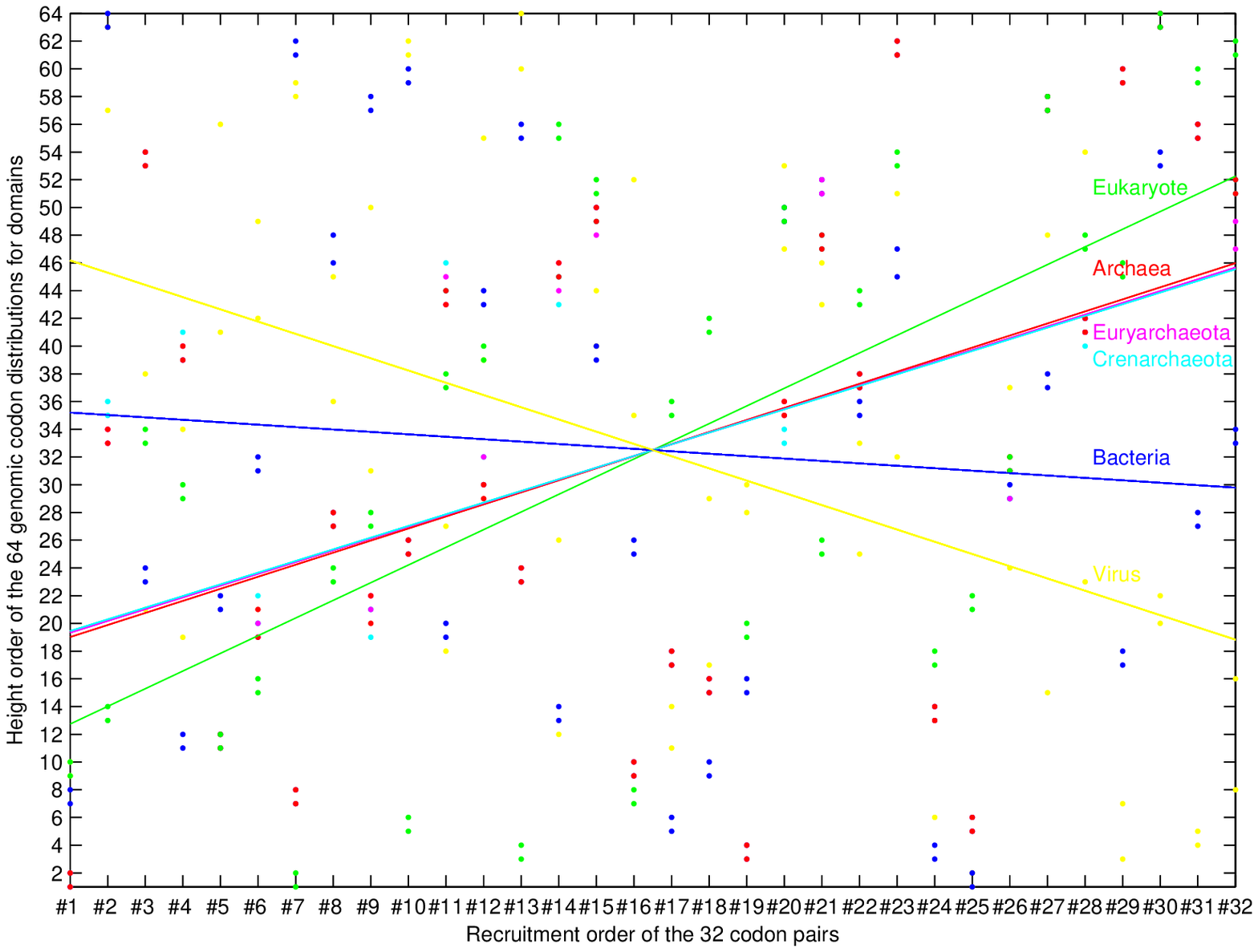}\\
  {\small \bf d} \includegraphics[width=16.5cm]{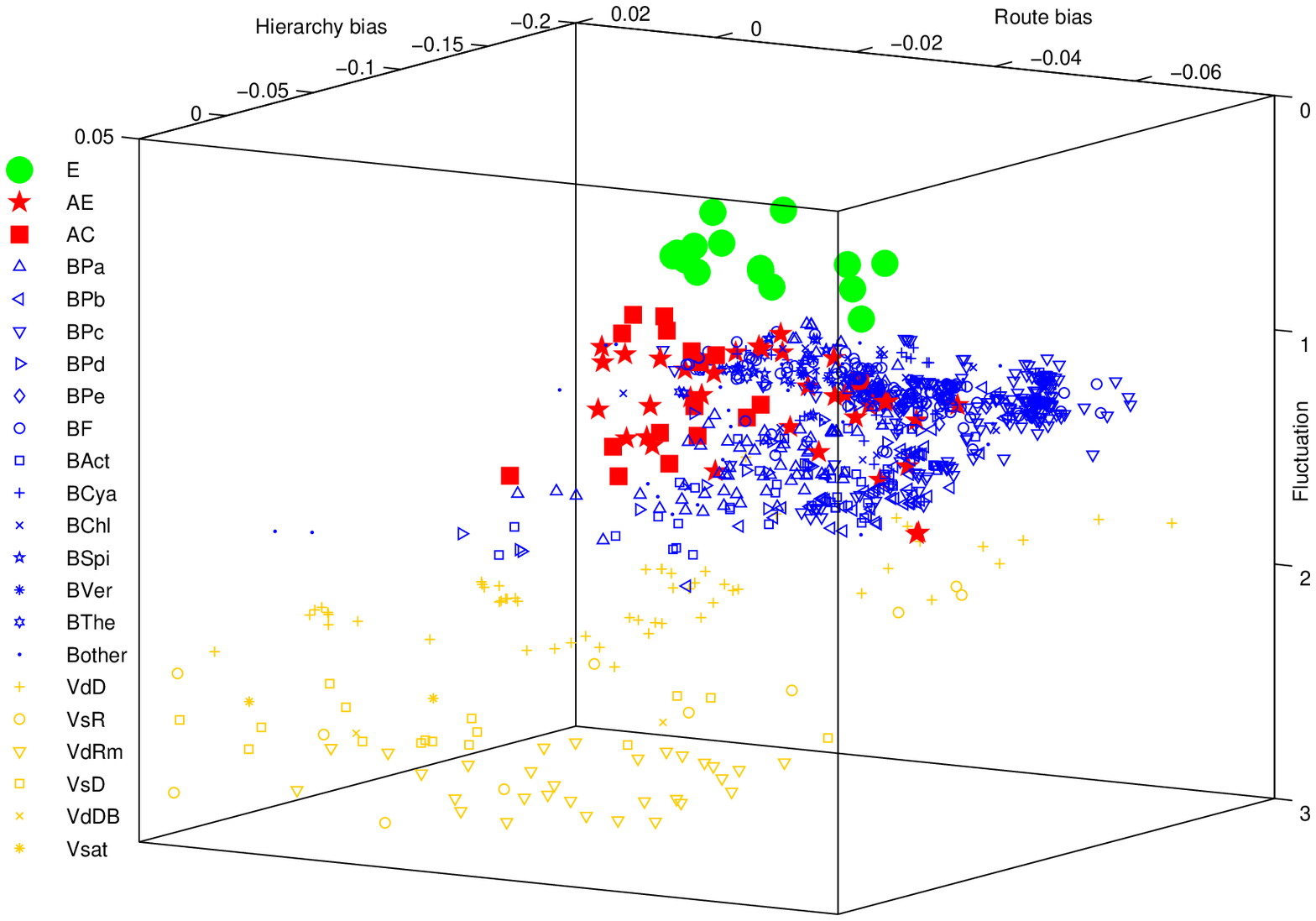}
\end{figure}

\clearpage \begin{figure}
  \centering
  \caption{Explanation of the origin of the three domains of life. {\bf a} The tree of codons based on the genomic codon distributions, which agrees with the grouping of codons in the four hierarchies $Hierarchy\ 1 \sim 4$ ($H1\sim 4$) on the roadmap. This tree of codons is plotted based on the distance matrix for codons by averaging the correlation coefficients of genomic codon distributions among the species. Both the tree of codons in this figure and the tree of species in Fig 4e are obtained based on a same set of genomic codon distributions. {\bf b} Features of genomic codon distributions for domains based on complete genome sequences. Refer to Fig 5a to see the genomic codon distributions of Crenarchaeota and Euryarchaeota; refer to Fig 5b to see the genomic codon distributions for species; and refer to Fig 5c to see the simulations of the genomic codon distributions. {\bf c} On the origin order of the domains. The recruitment order of the 32 codon pairs indicates an evolutionary direction, which reveals the origin order of the domains. {\bf d} Classification of Virus (yellow), Bacteria (blue), Archaea (red) and Eukarya (green) in the biodiversity space, which is based on the roadmap and the complete genome sequences. Refer to Fig 7c, 7d, 7e to see the two dimensional projections of this figure.}
\end{figure}

\clearpage \begin{figure}
 \centering
 {\small \bf a} \includegraphics[width=16.5cm]{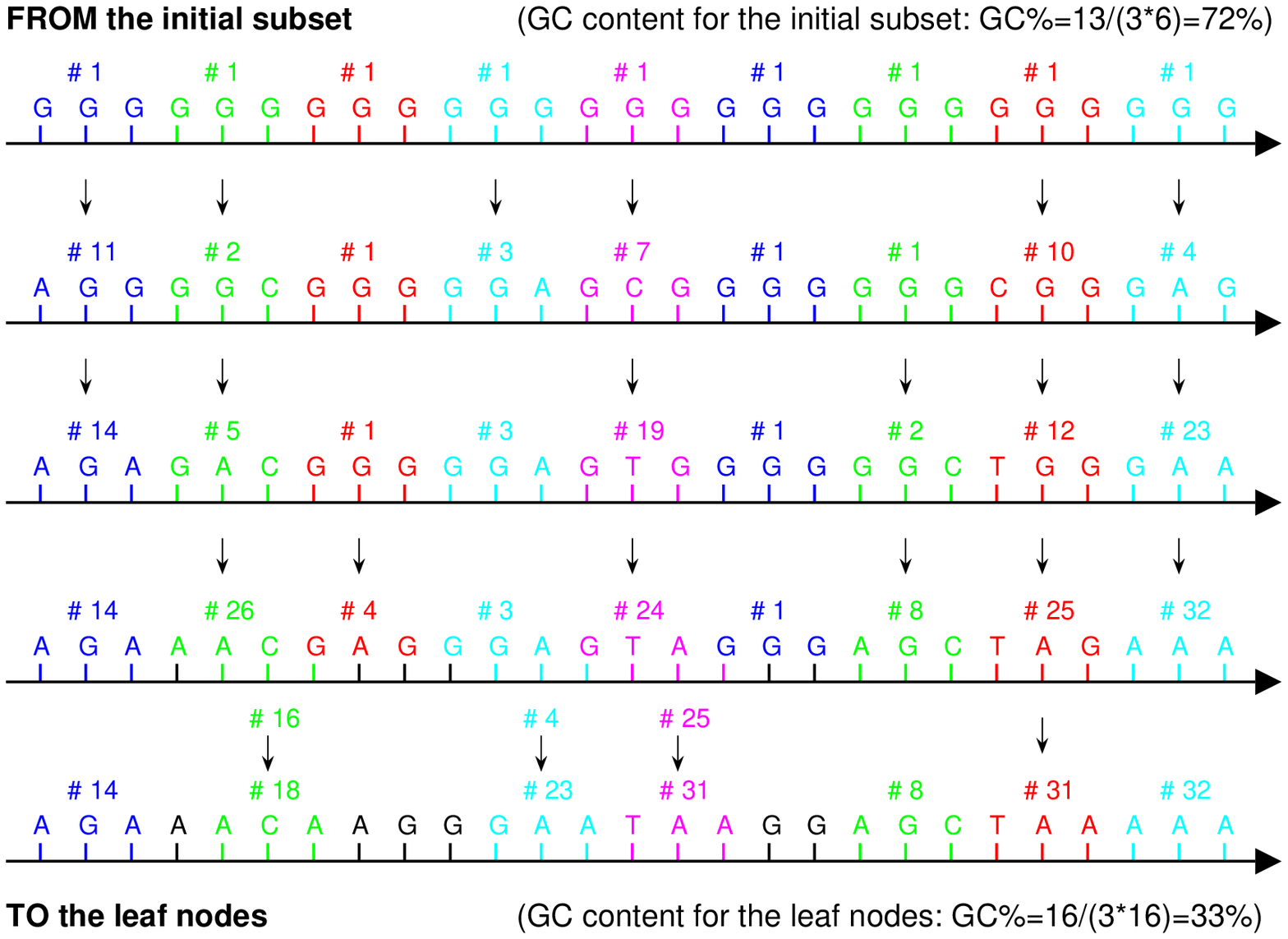}\\
 {\small \bf b} \includegraphics[width=16.5cm]{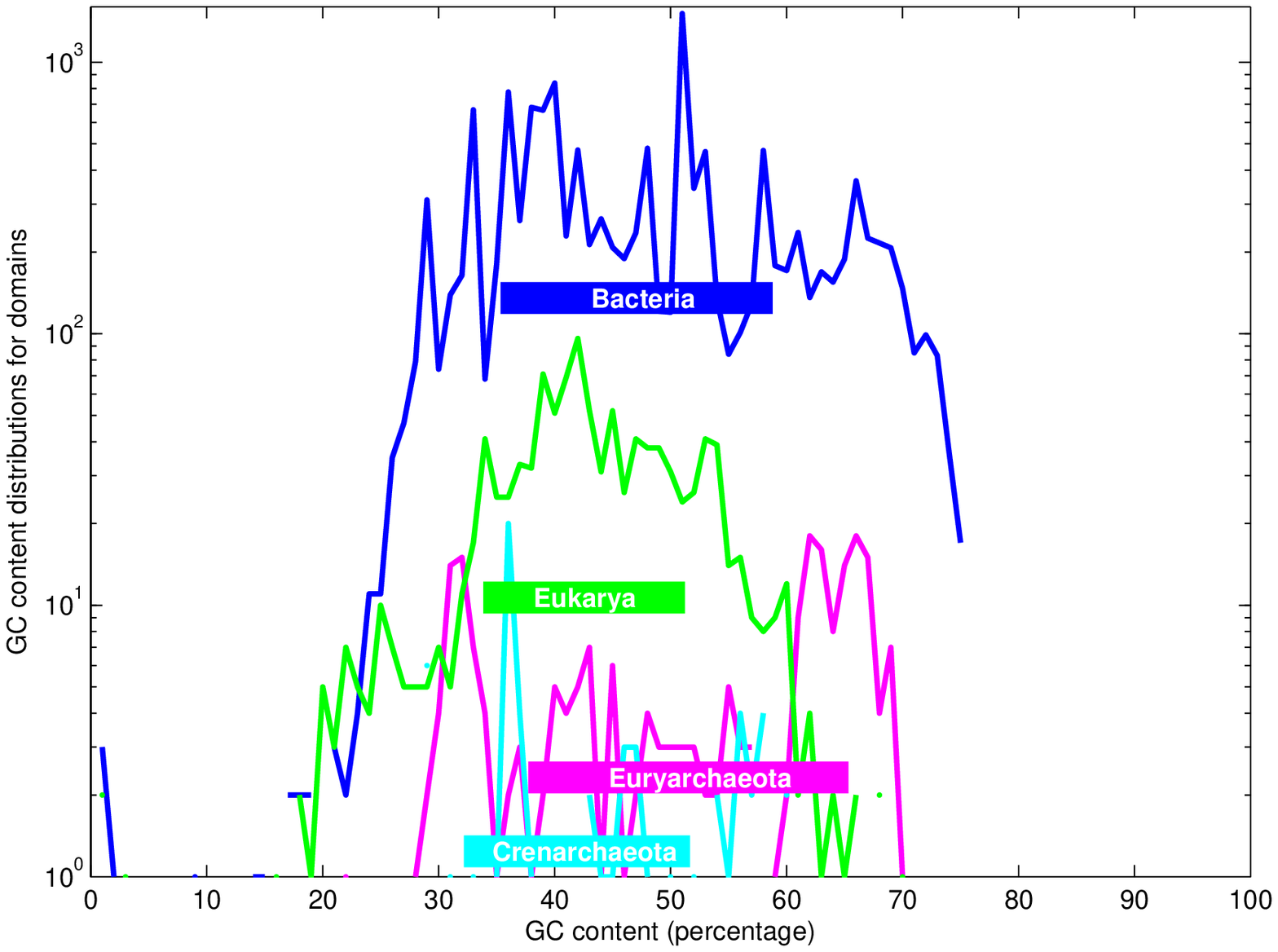}
\end{figure}

\clearpage \begin{figure}
 \centering
 {\small \bf c} \includegraphics[width=16.3cm]{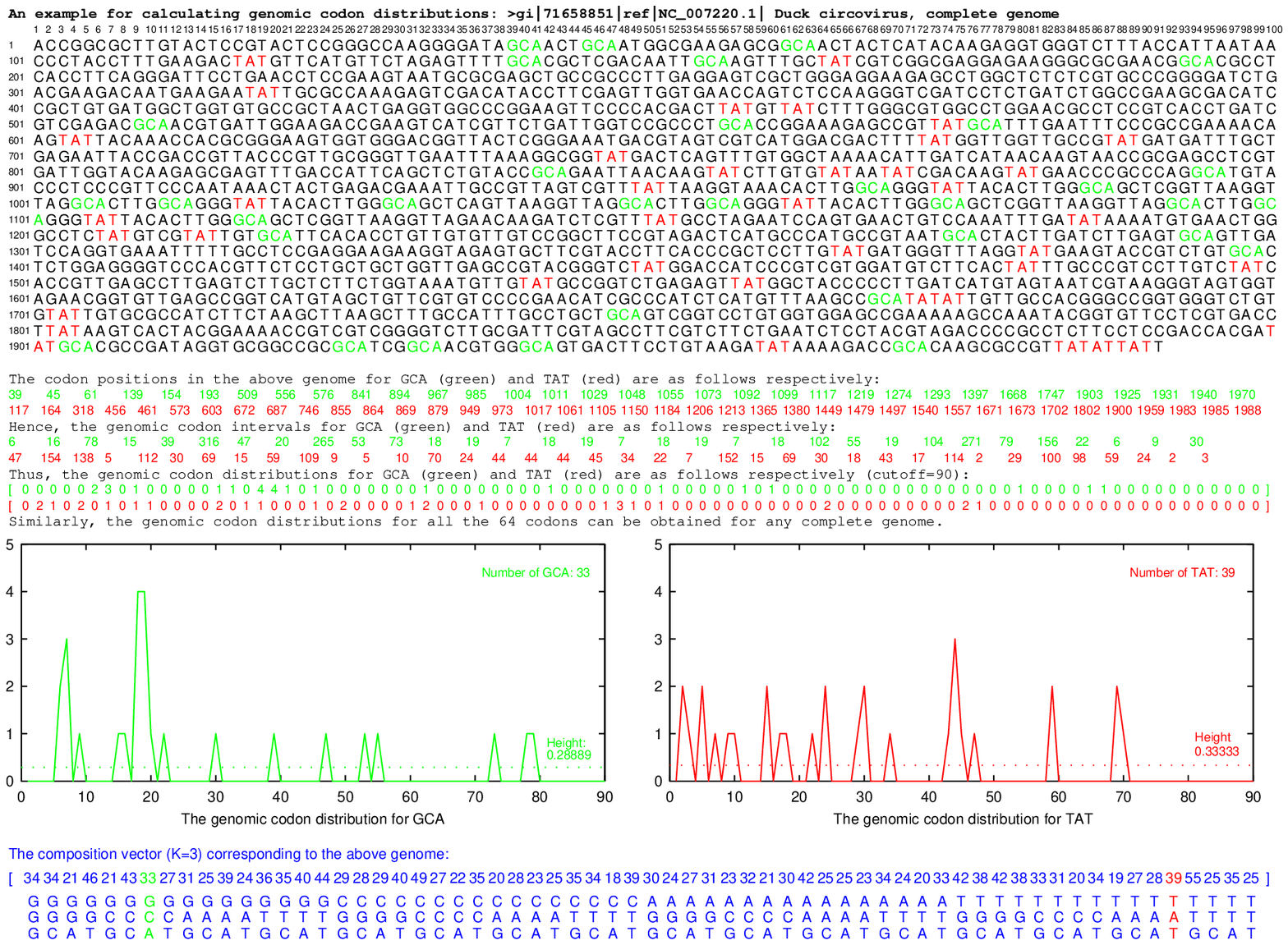}\\ \vskip 0.5cm
 {\small \bf d} \includegraphics[width=16.3cm]{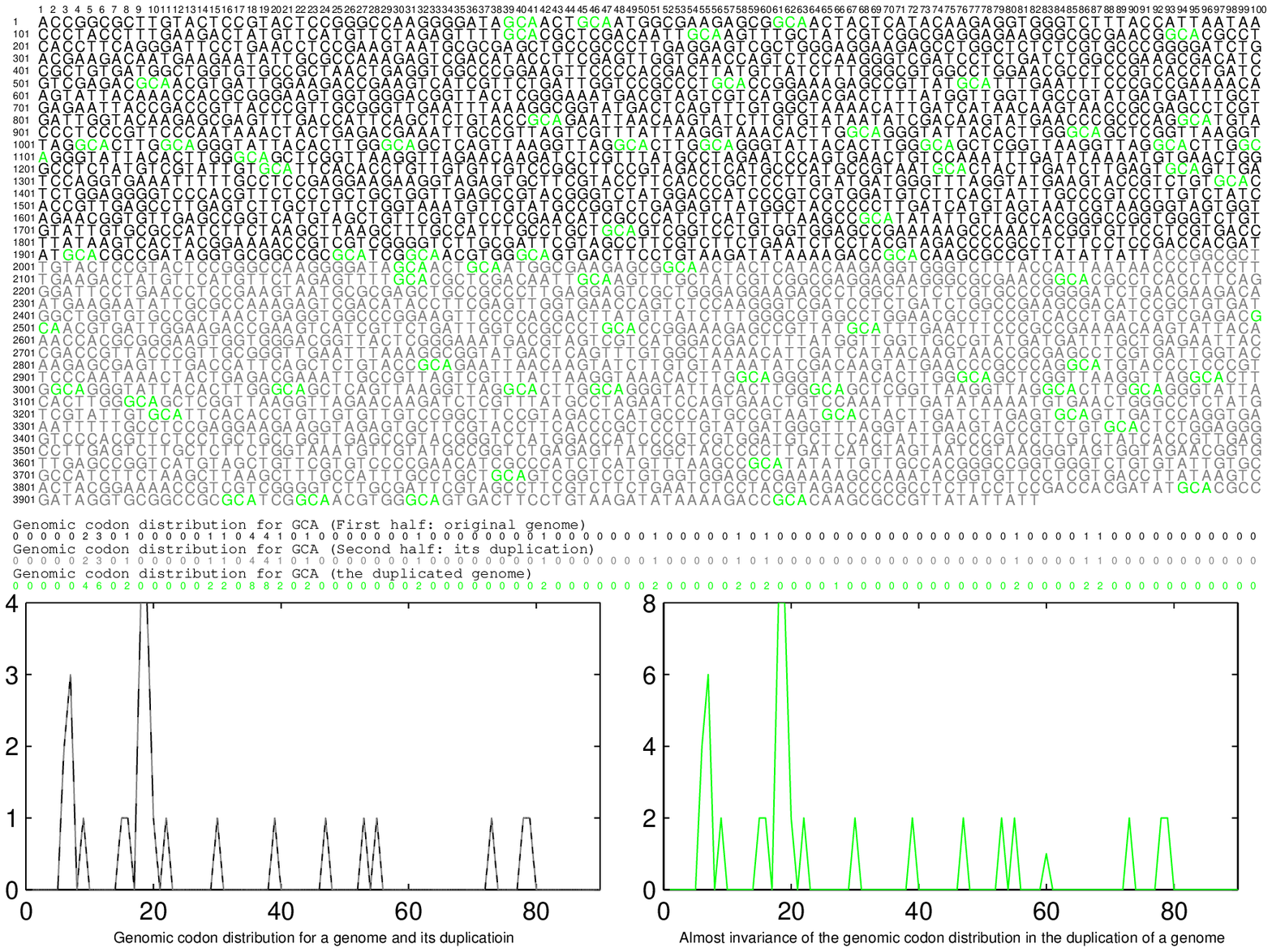}
\end{figure}

\clearpage \begin{figure}
 \centering
 \includegraphics[width=18.3cm]{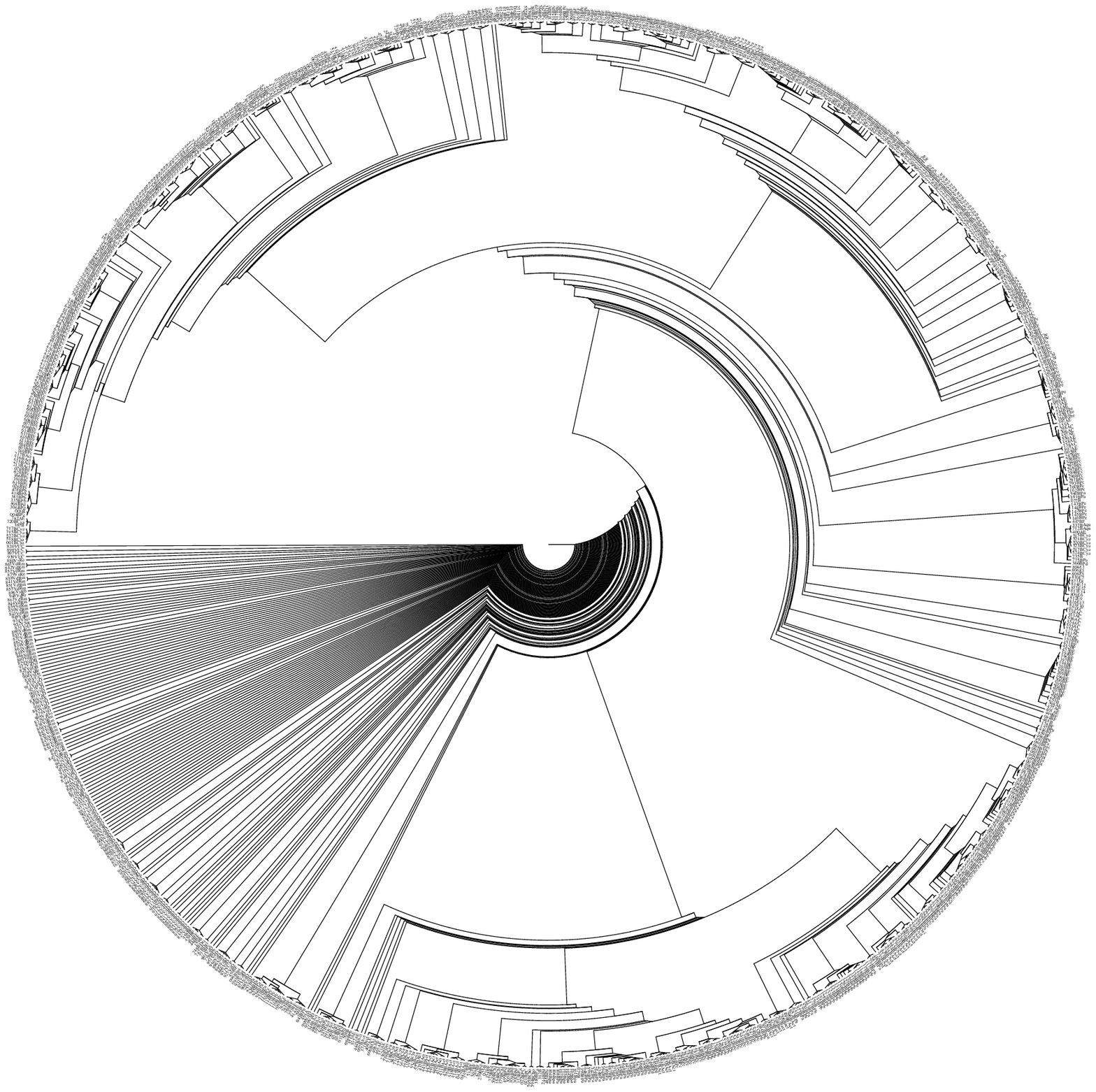}\\
 \vspace{1cm} {\small \bf e} 
\end{figure}

\clearpage \begin{figure}
 \centering
 \includegraphics[width=9cm]{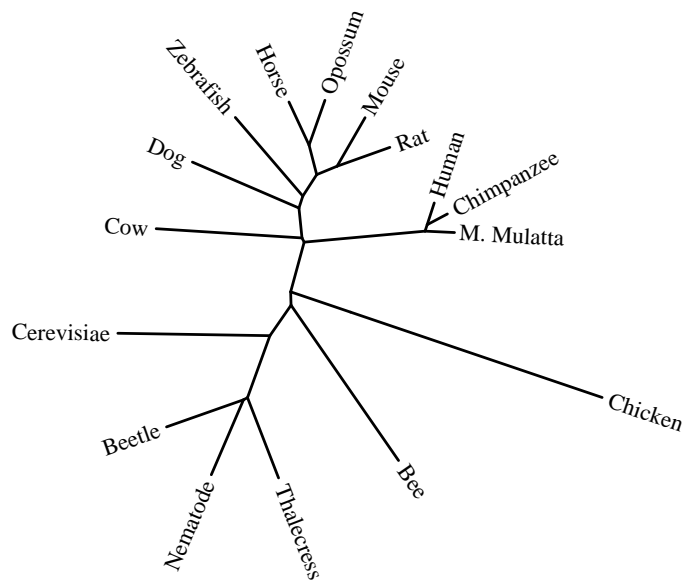}\\
 \vspace{-1.5cm} {\small \bf f} 
 \caption{Properties of genomic codon distributions. {\bf a} The sequence evolution starting from $Poly\ G$ based on the base substitutions along the roadmap. {\bf b} The wide ranges of $GC$ content for taxa in observations can be explained by the roadmap that the $GC$ content is about $72\%$ for the initial subset and about $33\%$ for the leaf nodes. {\bf c} Definition of the genomic codon distribution. Take the complete genome of Duck circovirus and the codons $GCA$ (denoted in green) and $TAT$ (denoted in red) for example. {\bf d} Additivity of the genomic codon distribution after genome duplication. {\bf e} Tree of species according to the average correlation coefficients among genomic codon distributions by averaging for the $64$ codons (enlarge to see details). Refer to the legend of Fig 3d to see the abbreviations of the taxa. {\bf f} Tree of eukaryotes based on the average correlation coefficients among genomic codon distributions. This tree is reasonable to some extent.} 
\end{figure}

\clearpage  \begin{figure}
 \centering
 {\small \bf a} \includegraphics[width=17.8cm]{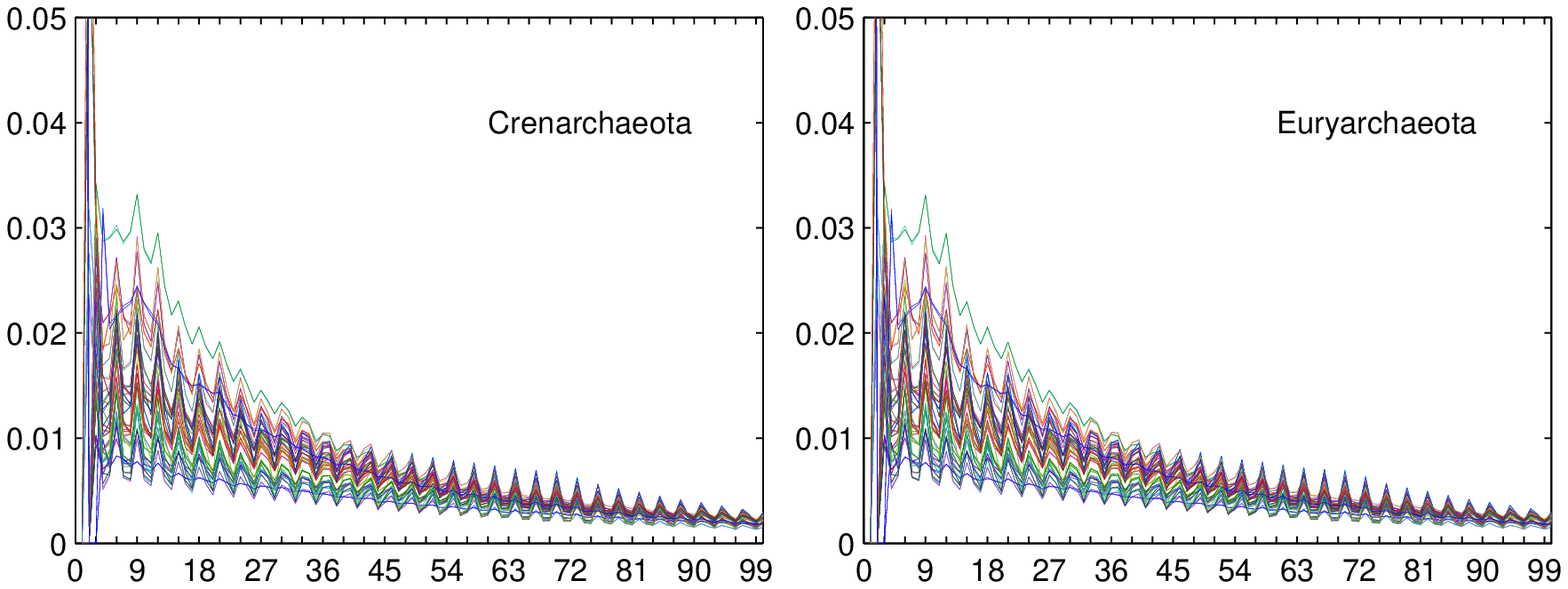}\\ \vskip 2cm
 {\small \bf b} \includegraphics[width=17.8cm]{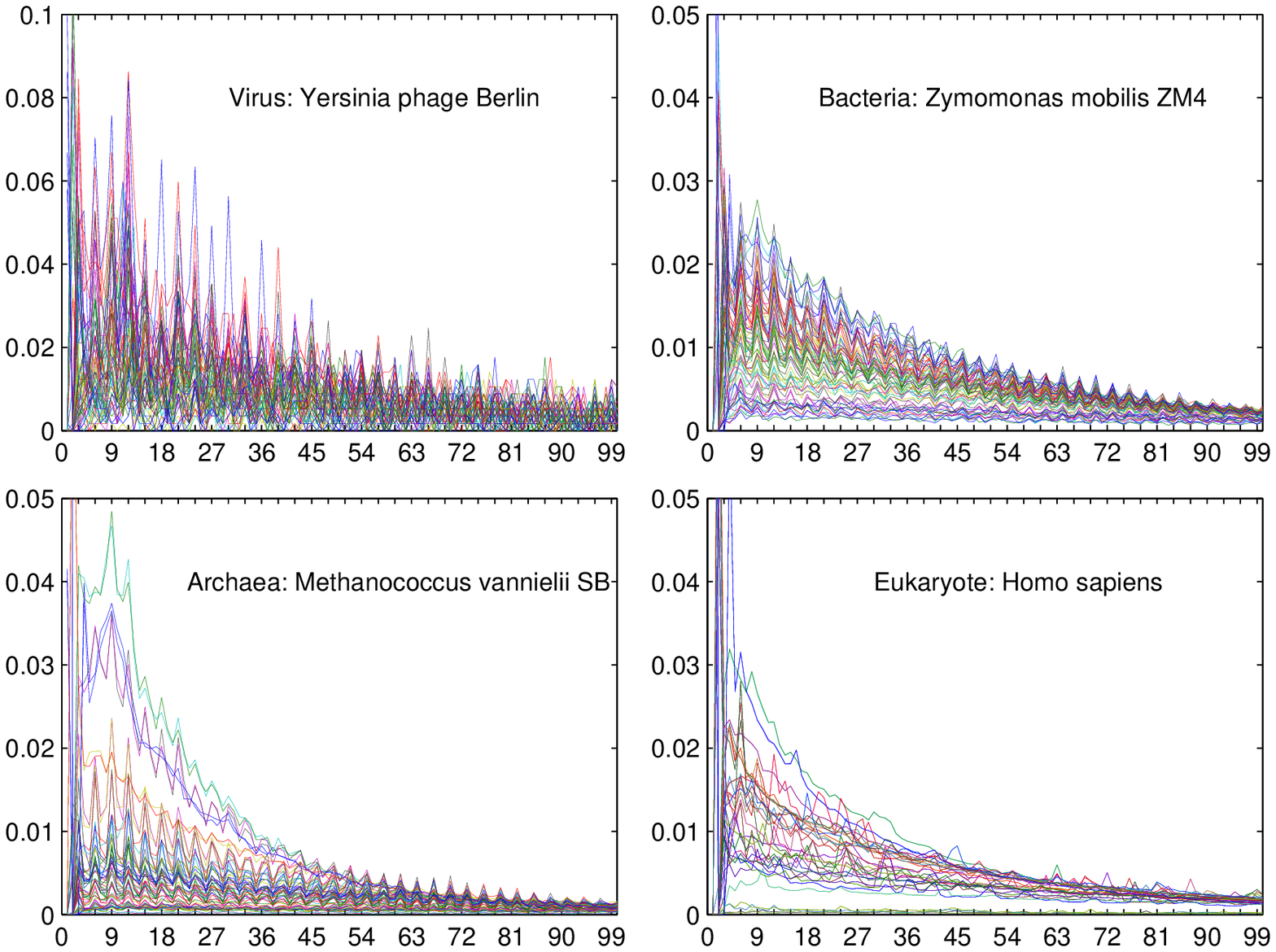}
 \end{figure}

\clearpage  \begin{figure}
 \centering
 {\small \bf c} \includegraphics[width=17.8cm]{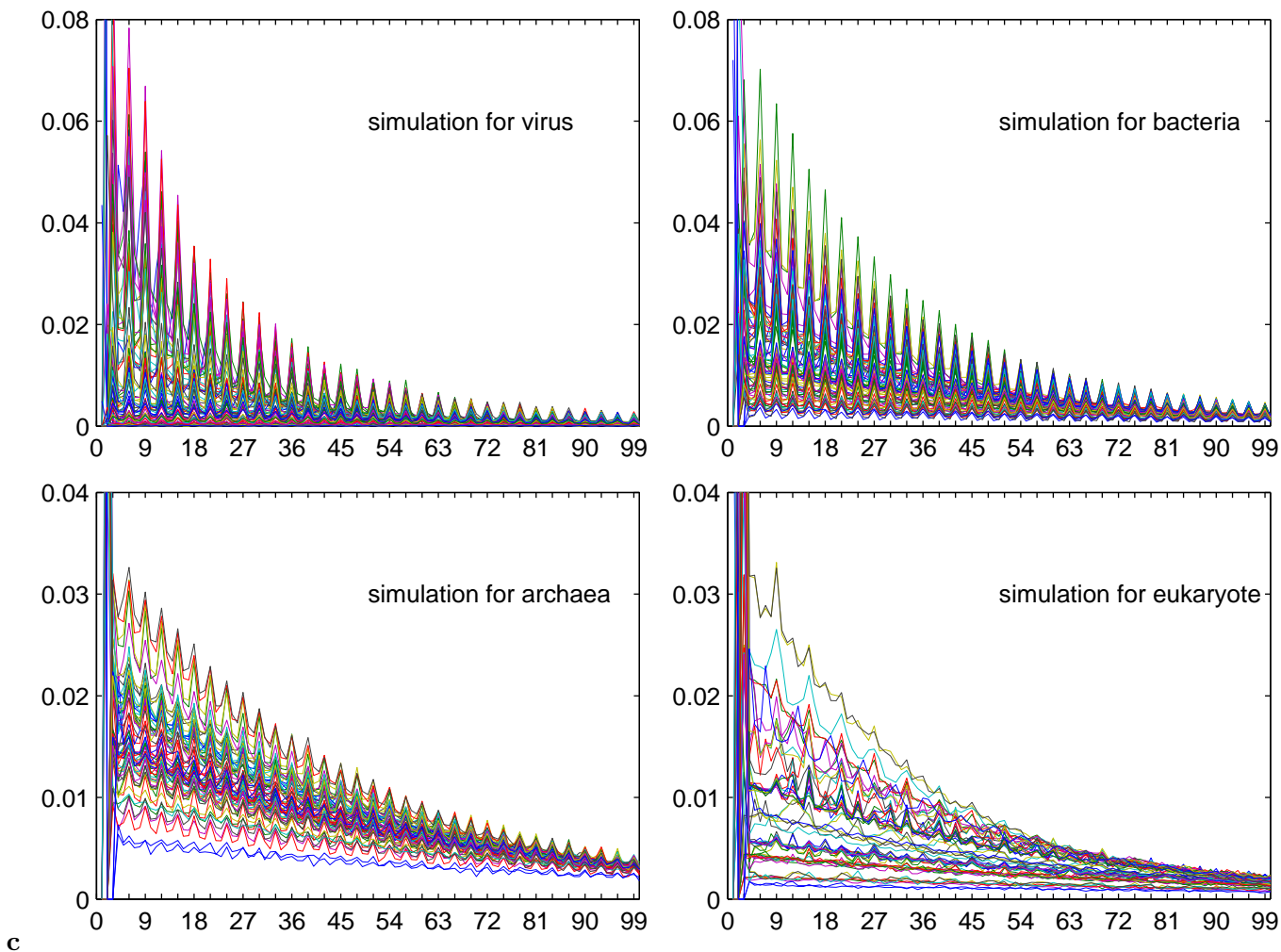}
 \caption{Supplements to Fig 1b. {\bf a} Genomic codon distributions for Crenarchaeota and Euryarchaeota based on genomic data. The features for both Crenarchaeota and Euryarchaeota agree with that for Archaea in Fig 3b, while they are quite different from the features for Eukarya in Fig 3b. {\bf b} Genomic codon distributions for species. The features for species agree respectively with the features for the domains in Fig 3b. {\bf c} Simulations of the features of genomic codon distributions for the different domains. This simulation results for the domains agree with the observations based on genomic data.}
\end{figure}

\clearpage \begin{figure}
 \centering
 {\small \bf a} \includegraphics[width=16cm]{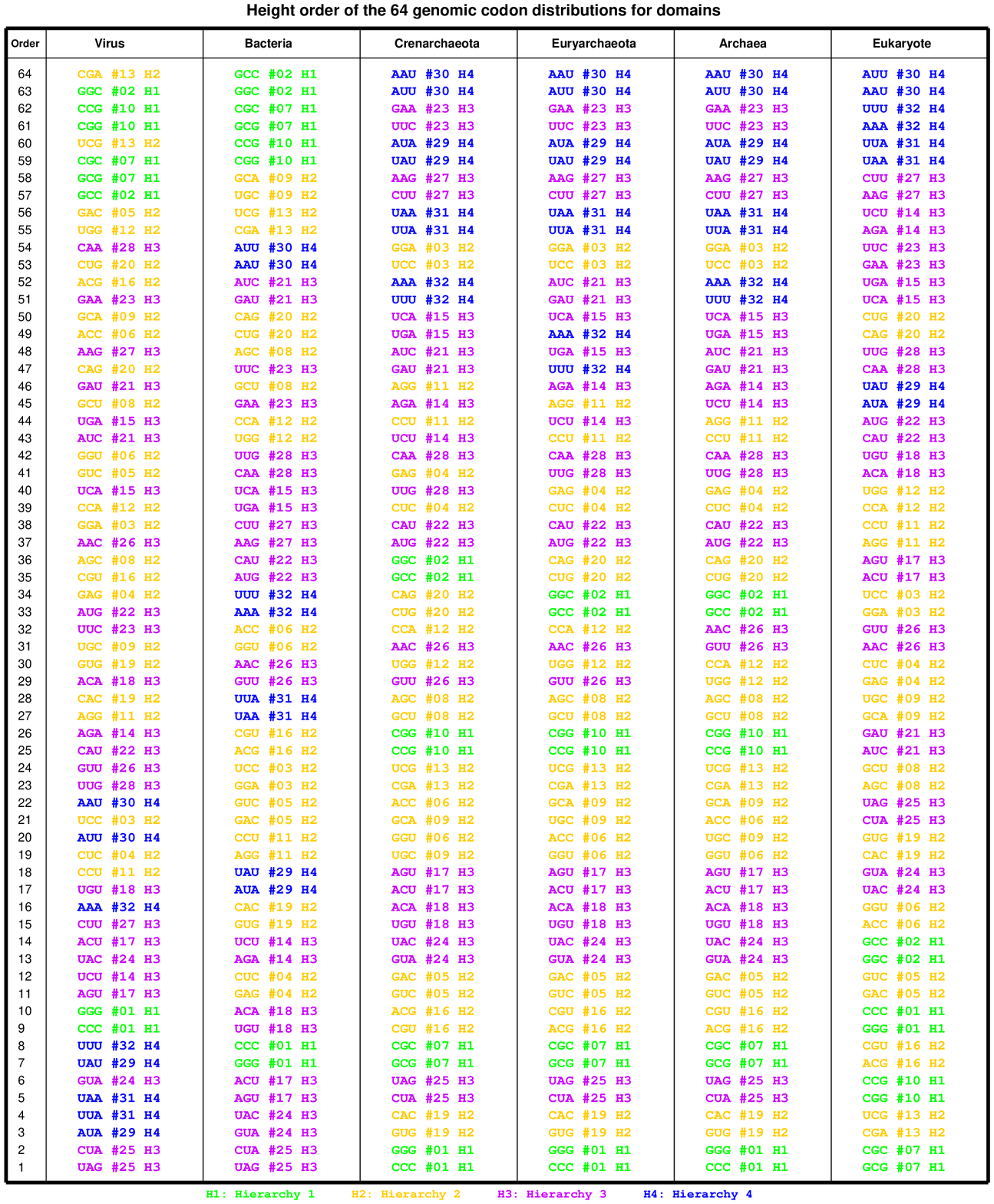}
 \end{figure}
 
\clearpage \begin{figure}
 \centering
 {\small \bf b} \includegraphics[width=16cm]{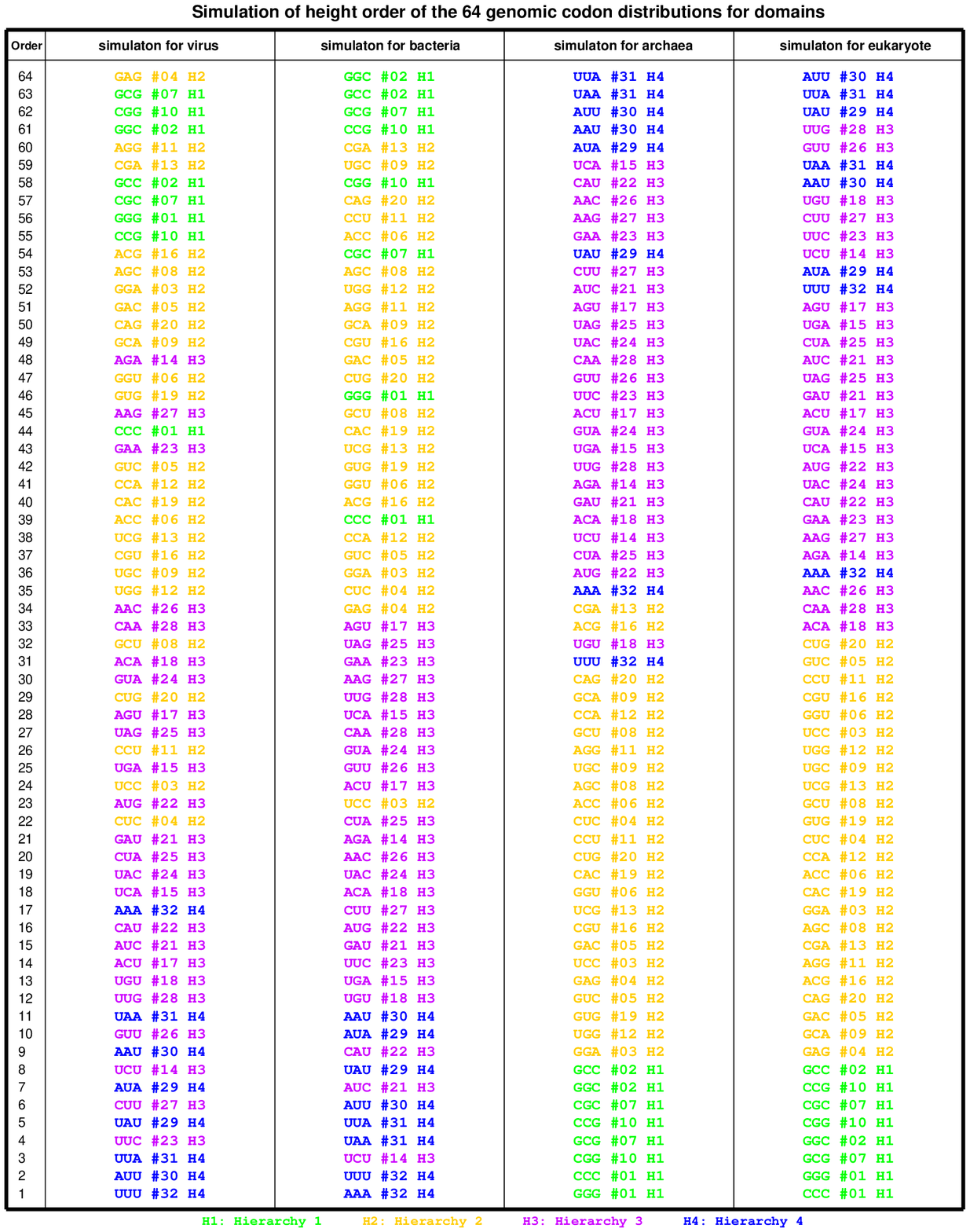}
 \caption{{\bf a} The orders of the average distribution heights for different domains based on the genomic data that are also used in Fig 3b and Fig 5a. {\bf b} Simulations of the orders of the average distribution heights for different domains, where the values of genomic codon distributions here are obtained from the simulation results in Fig 5c. The simulation results agree with the observations in Fig 6a based on genomic data.}
\end{figure}

\clearpage \begin{figure}
 \centering
 {\small \bf a} \includegraphics[width=12cm]{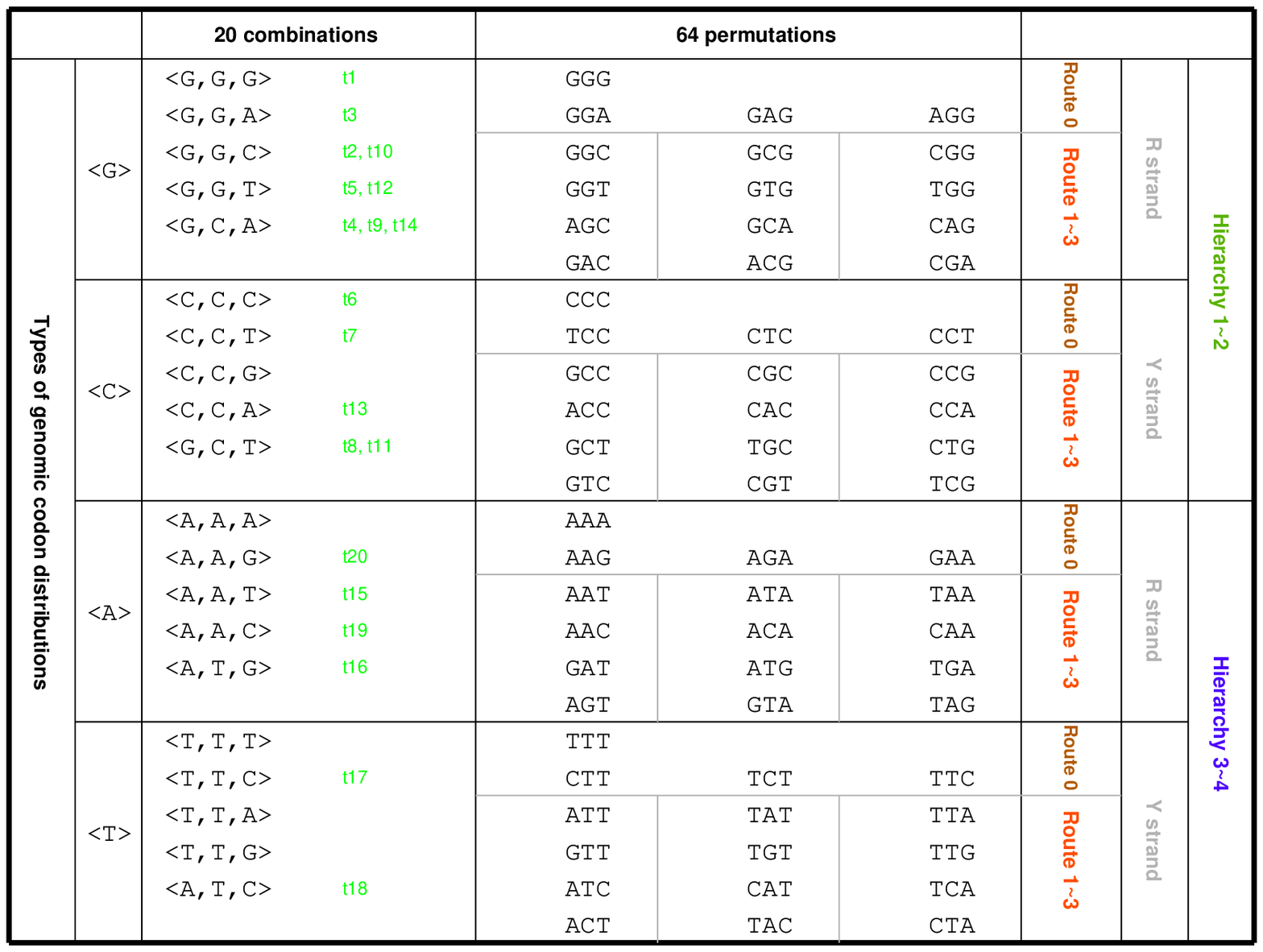}\\
 {\small \bf b} \includegraphics[width=12cm]{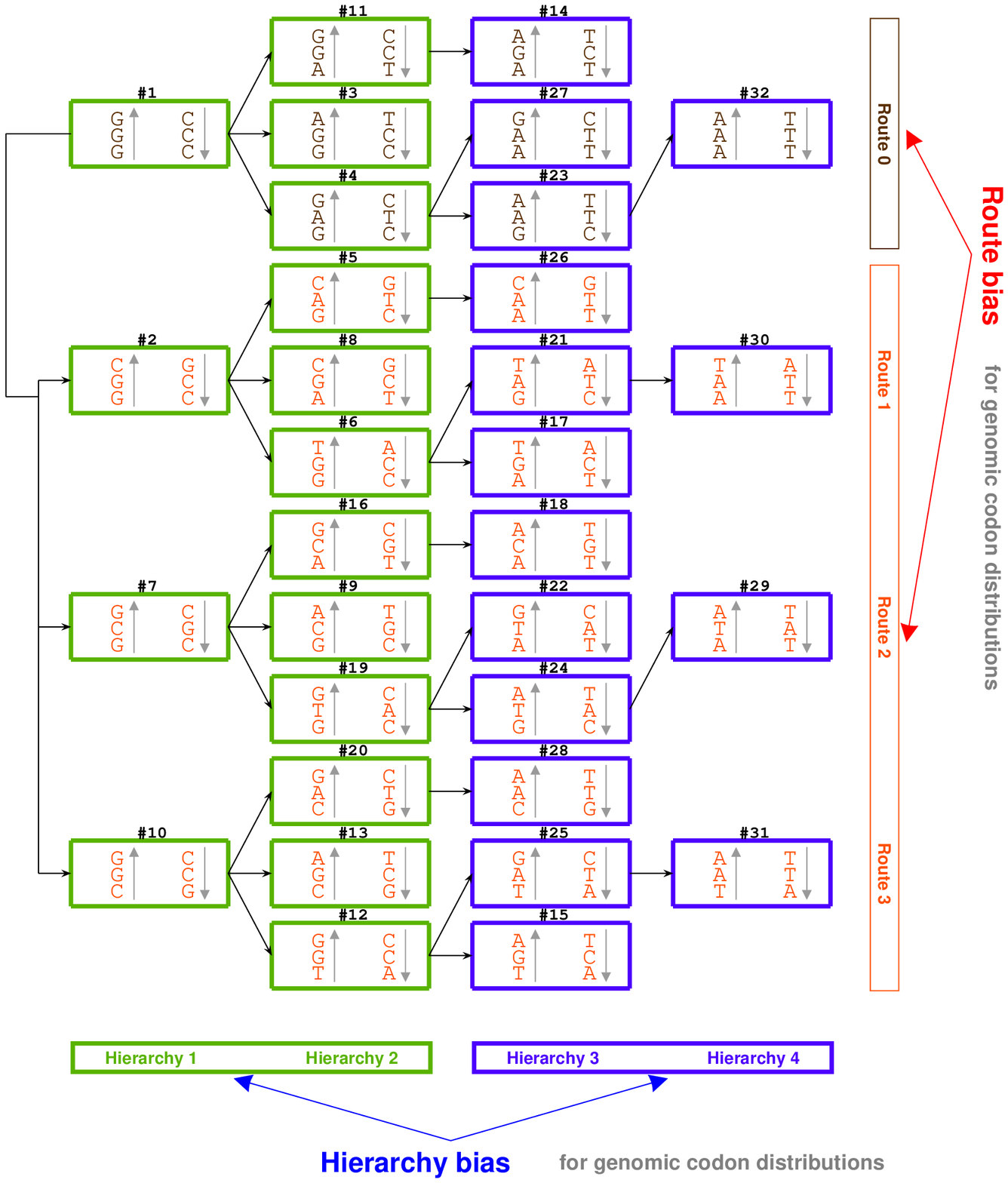}
\end{figure}

\clearpage \begin{figure}
 \centering
 {\small \bf c} \includegraphics[width=15.5cm]{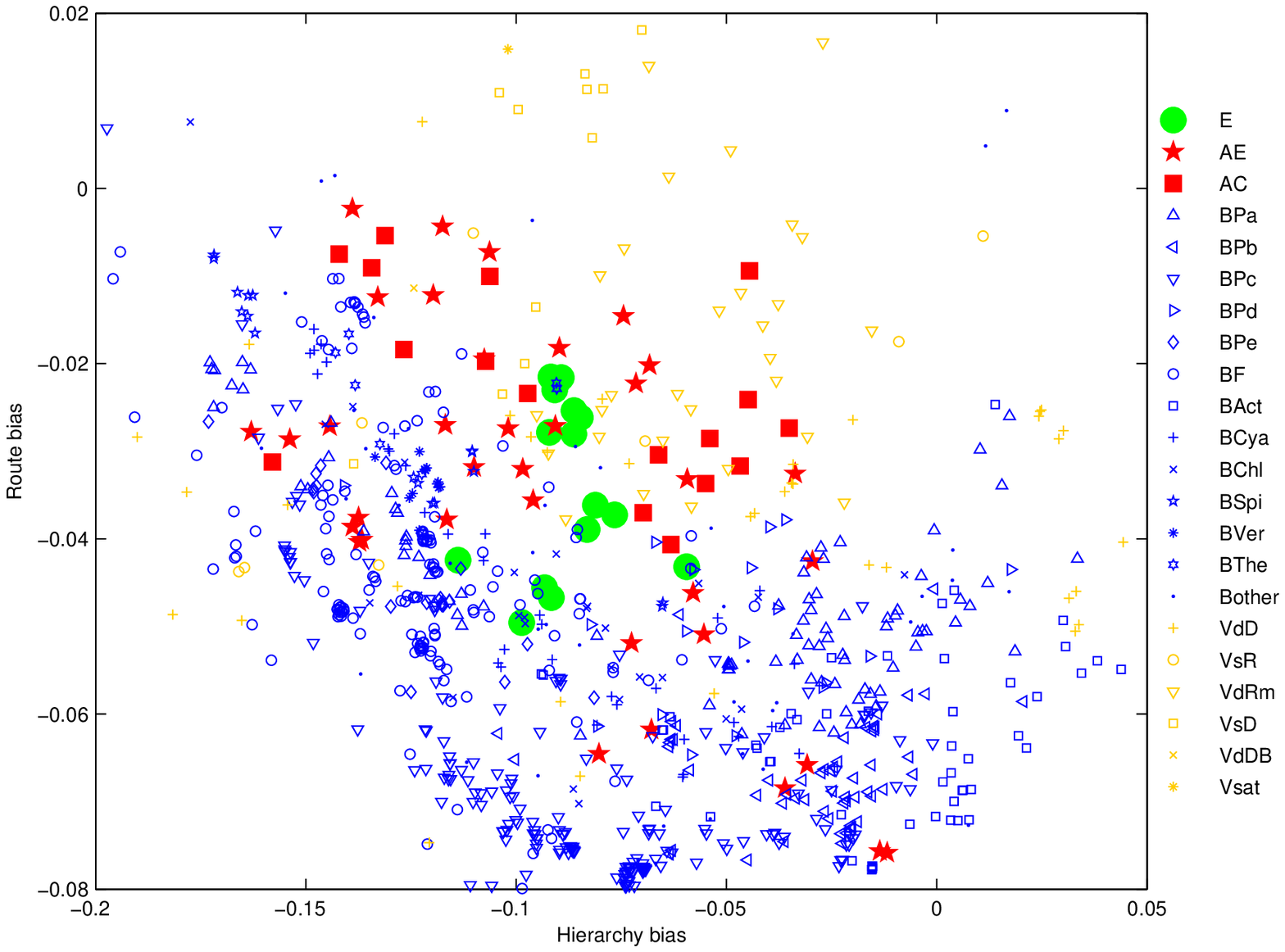}\\ \vskip 0.5cm
 {\small \bf d} \includegraphics[width=8.5cm]{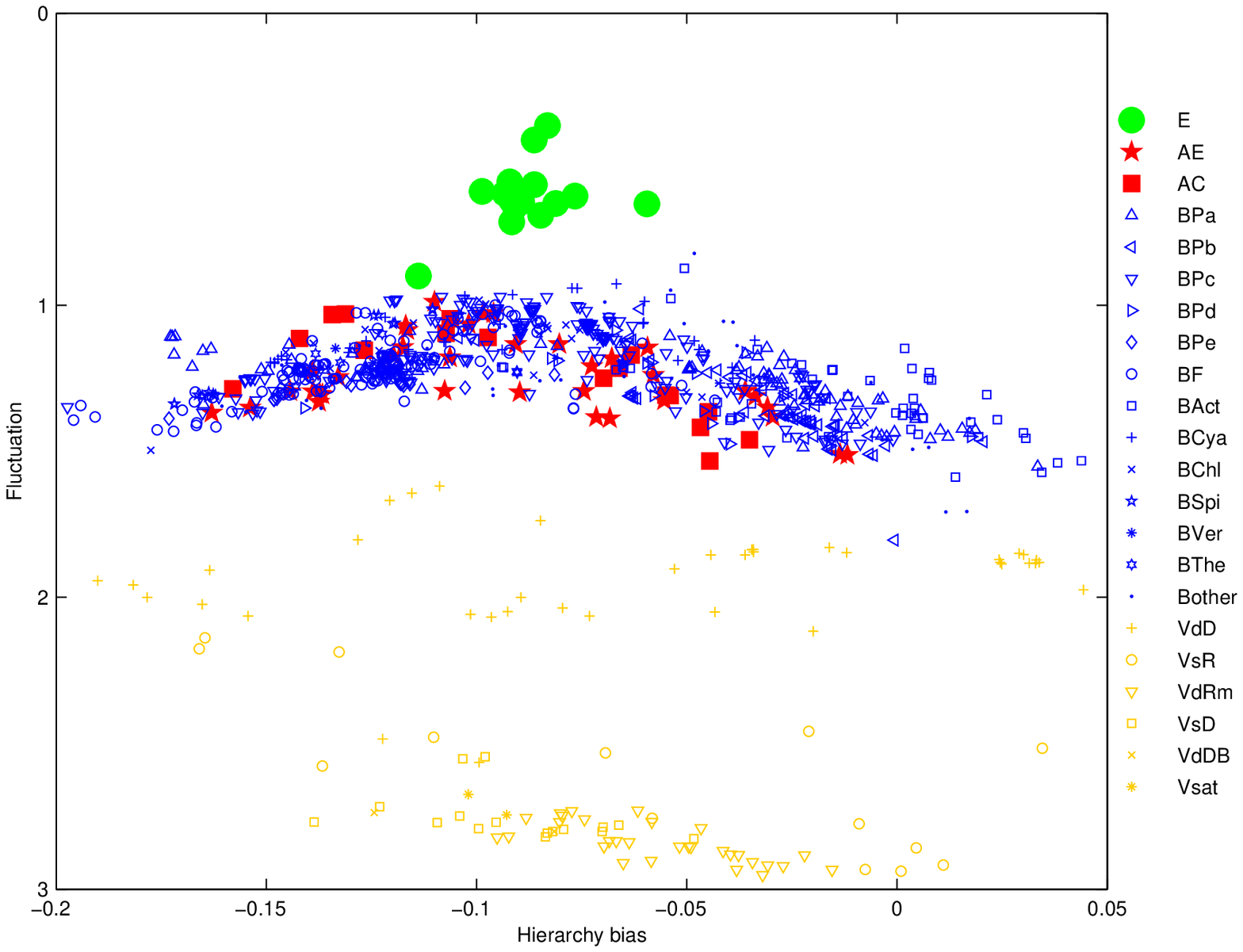}
 {\small \bf e} \includegraphics[width=8.5cm]{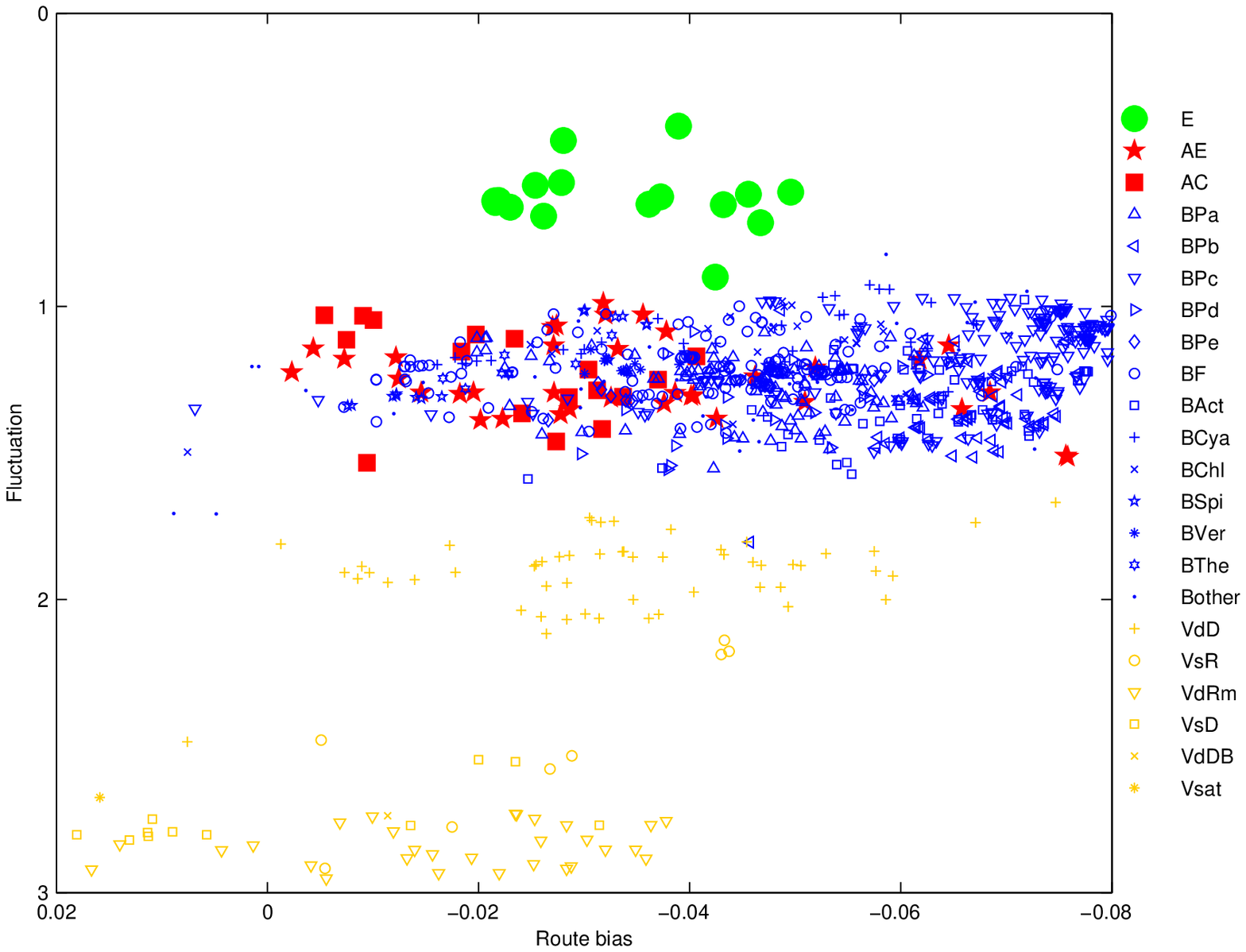}\\
 {\small \bf f} \includegraphics[width=8.5cm]{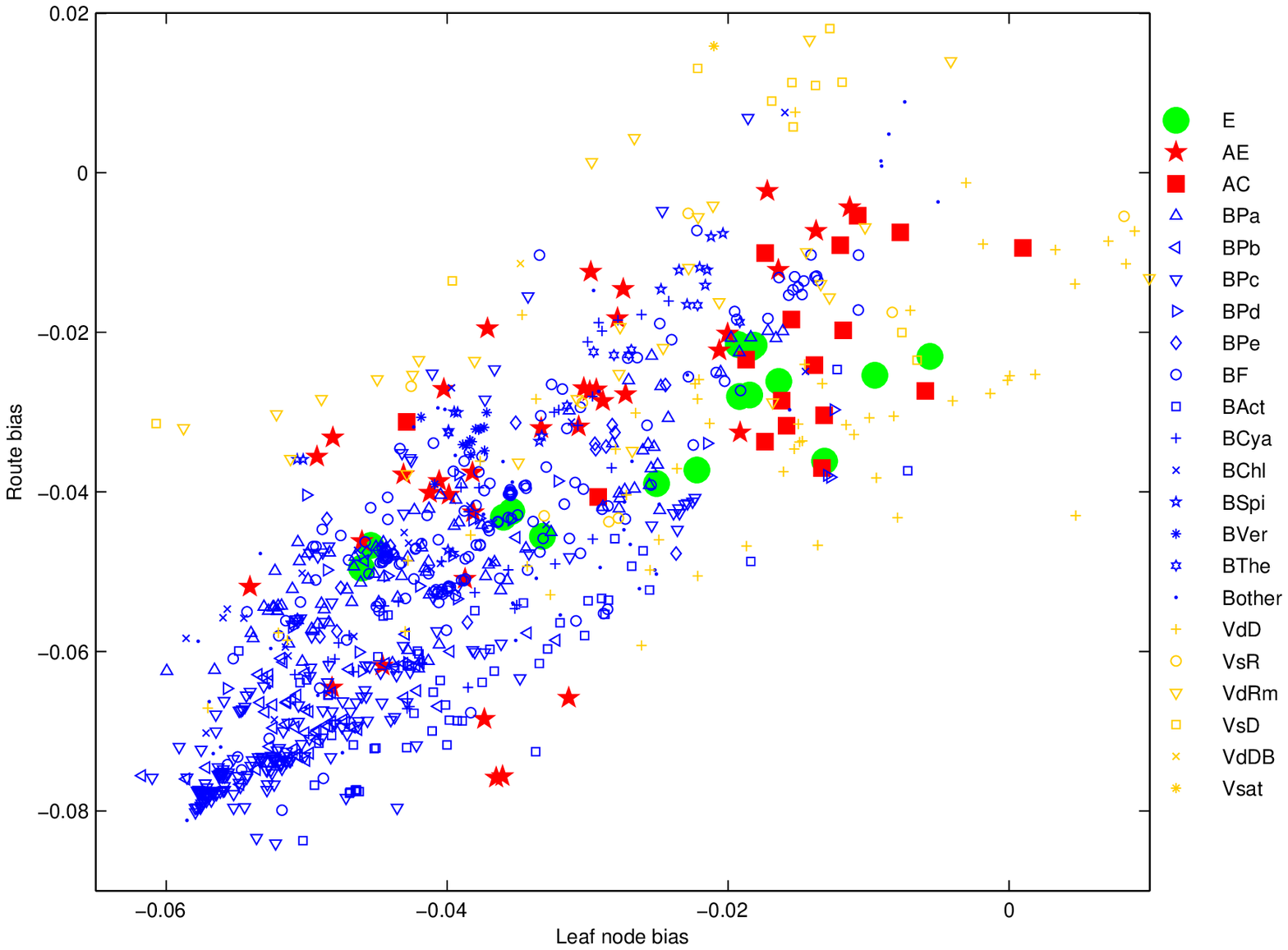}
 {\small \bf g} \includegraphics[width=8.5cm]{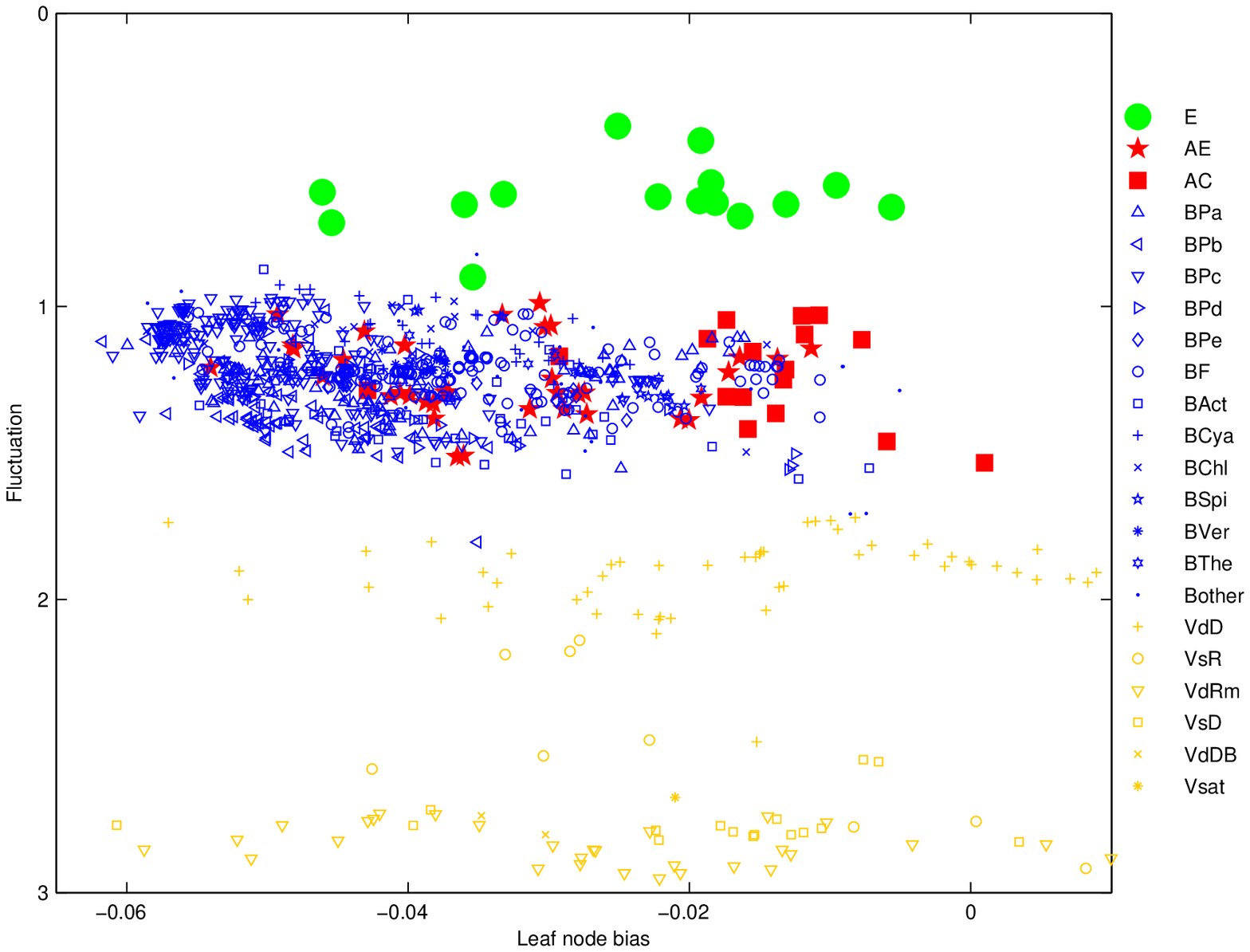}
\end{figure}

\clearpage \begin{figure}
 \centering
 \includegraphics[width=18.3cm]{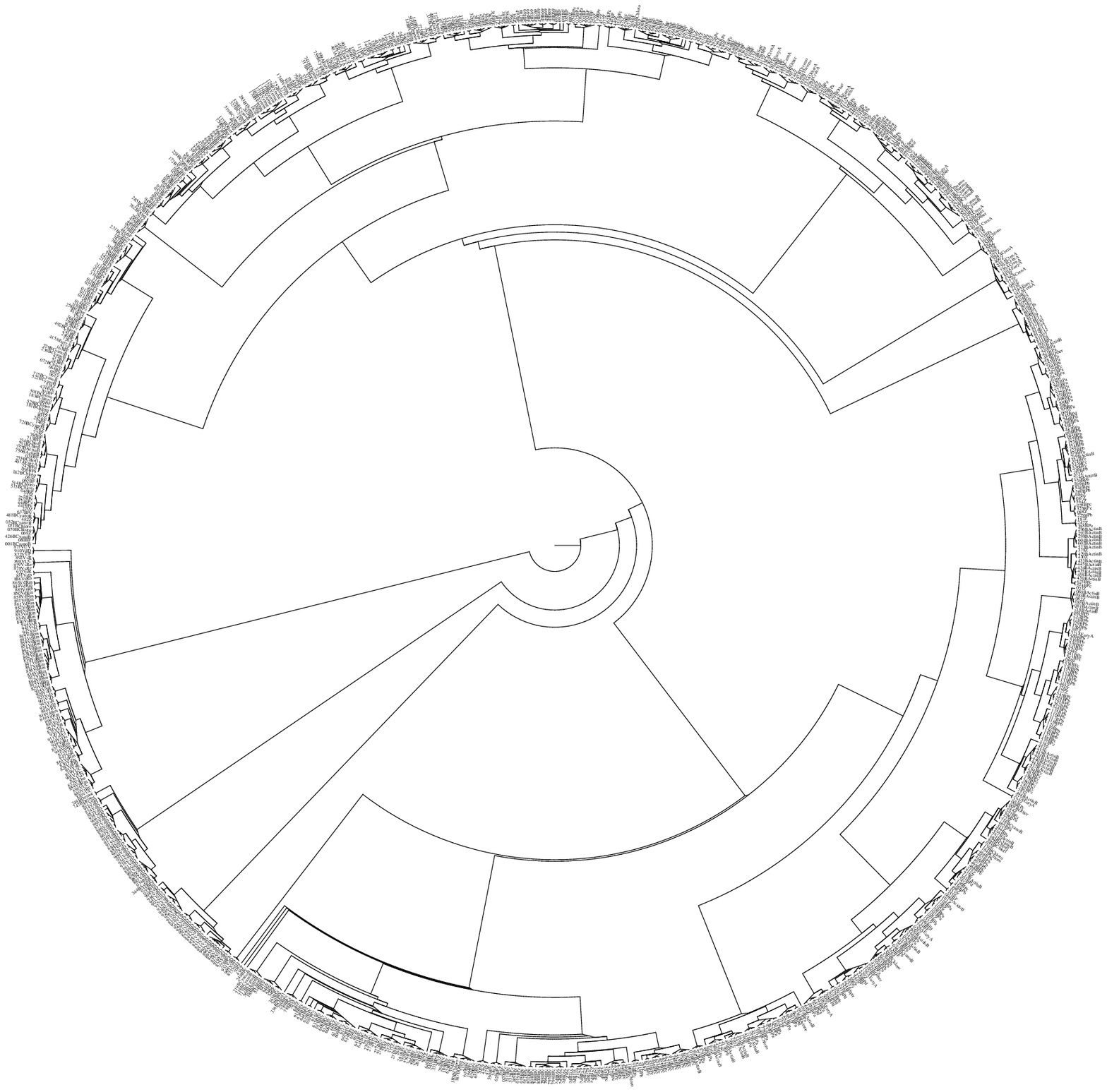}\\
 \vspace{1cm}{\small \bf h} 
\end{figure}

\clearpage \begin{figure}
 \centering
 {\small \bf i} \includegraphics[width=9cm]{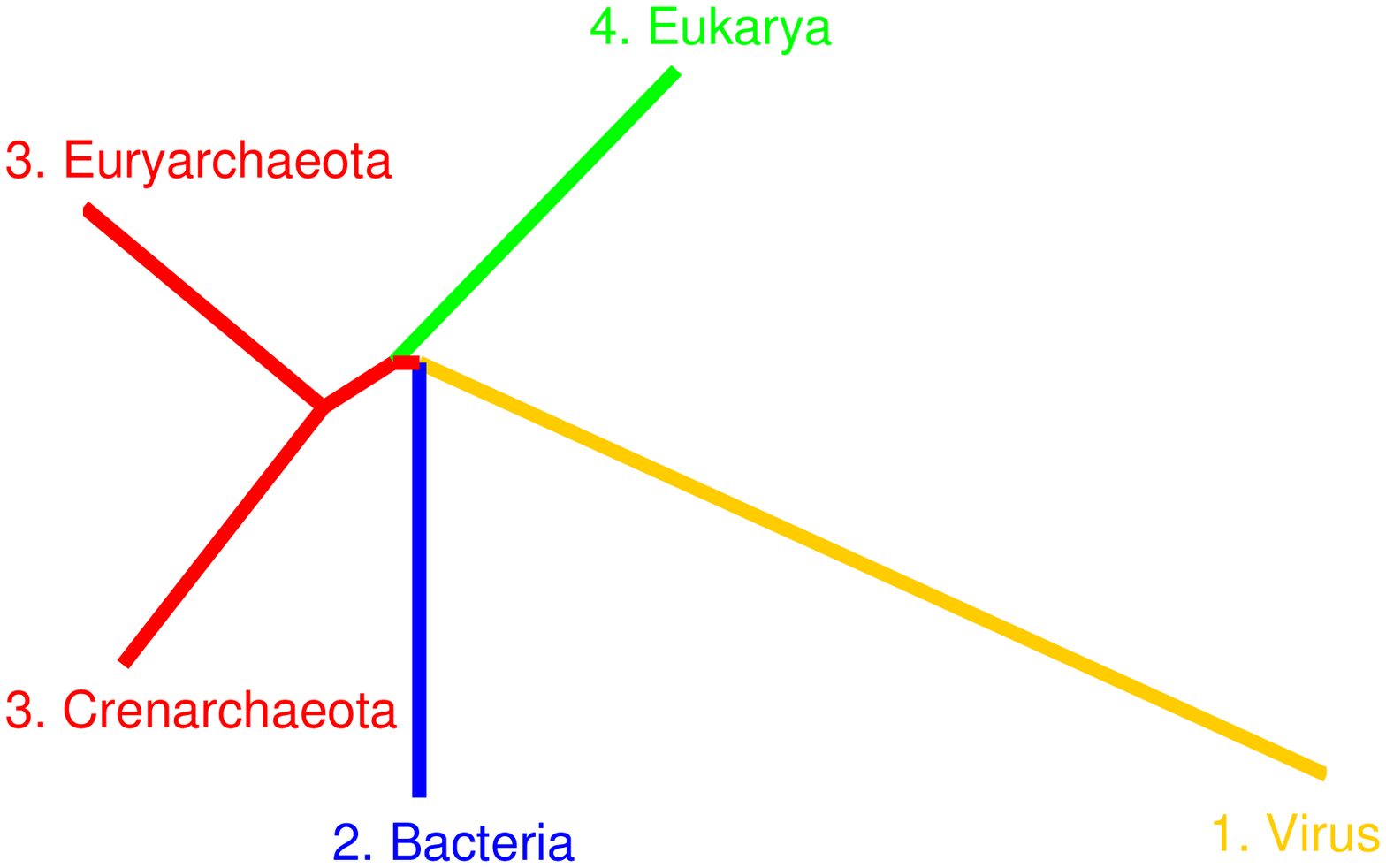}
 \hspace{2cm}{\small \bf j} \includegraphics[width=5cm]{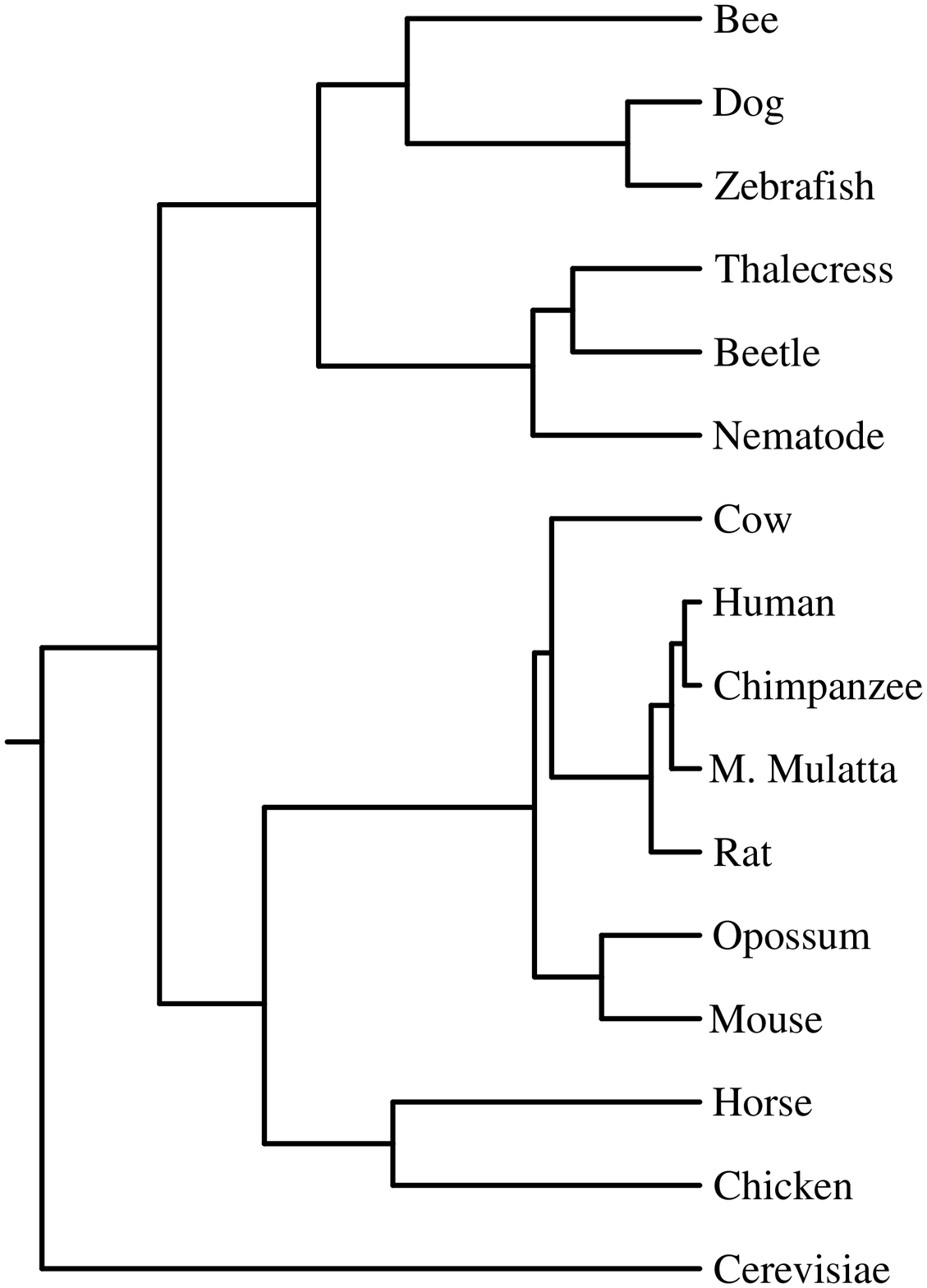}
 \caption{Explanation of the origin of the three domains based on the roadmap of genetic code evolution (Fig 1a in Li 2018-I). {\bf a} The $20$ types of genomic codon distributions. There are $20$ combinations and $64$ permutations for the $4$ bases. There is a coarse correlation between the $20$ combinations of codons and the $20$ amino acids according to the tRNAs $t1$ to $t20$ (Fig 6b in Li 2018-I). {\bf b} Explanation of the definitions of hierarchy bias and route bias. {\bf c} The $hierarchy\ bias$ and $route\ bias$ plane projection of the biodiversity space. Bacteria (blue) and Archaea (red) are distinguished by and large. {\bf d} The $hierarchy\ bias$ and $fluctuation$ plane projection of the biodiversity space. Eukarya (green), Bacteria, Archaea, and Virus (yellow) are distinguished by and large. {\bf e} The $route\ bias$ and $fluctuation$ plane projection of the biodiversity space. Eukarya (green), Bacteria (blue), Archaea (red), and Virus (yellow) are distinguished by and large. {\bf f} Distinguish Crenarchaeota (AC) and Euryarchaeota (AE) in the $leaf\ node\ bias$ and $route\ bias$ plane. {\bf g} Distinguish Crenarchaeota (AC) and Euryarchaeota (AE) in the $leaf\ node\ bias$ and $fluctuation$ plane. {\bf h} Tree of life based on the Euclidean distances among species in the nondimentionalised biodiversity space Fig 3d. {\bf i} Tree of Bacteria, Crenarchaeota, Euryarchaeota, Eukarya and Virus based on the average distances among species in these taxa in the nondimentionalised biodiversity space Fig 3d. This tree agrees with the three-domain tree rather than the eocyte tree. {\bf j} Tree of eukaryotes based on their distances in the nondimentionalised biodiversity space Fig 3d.}
\end{figure}

\clearpage  
\begin{figure}
 \centering
 {\small \bf a} \includegraphics[width=8.5cm]{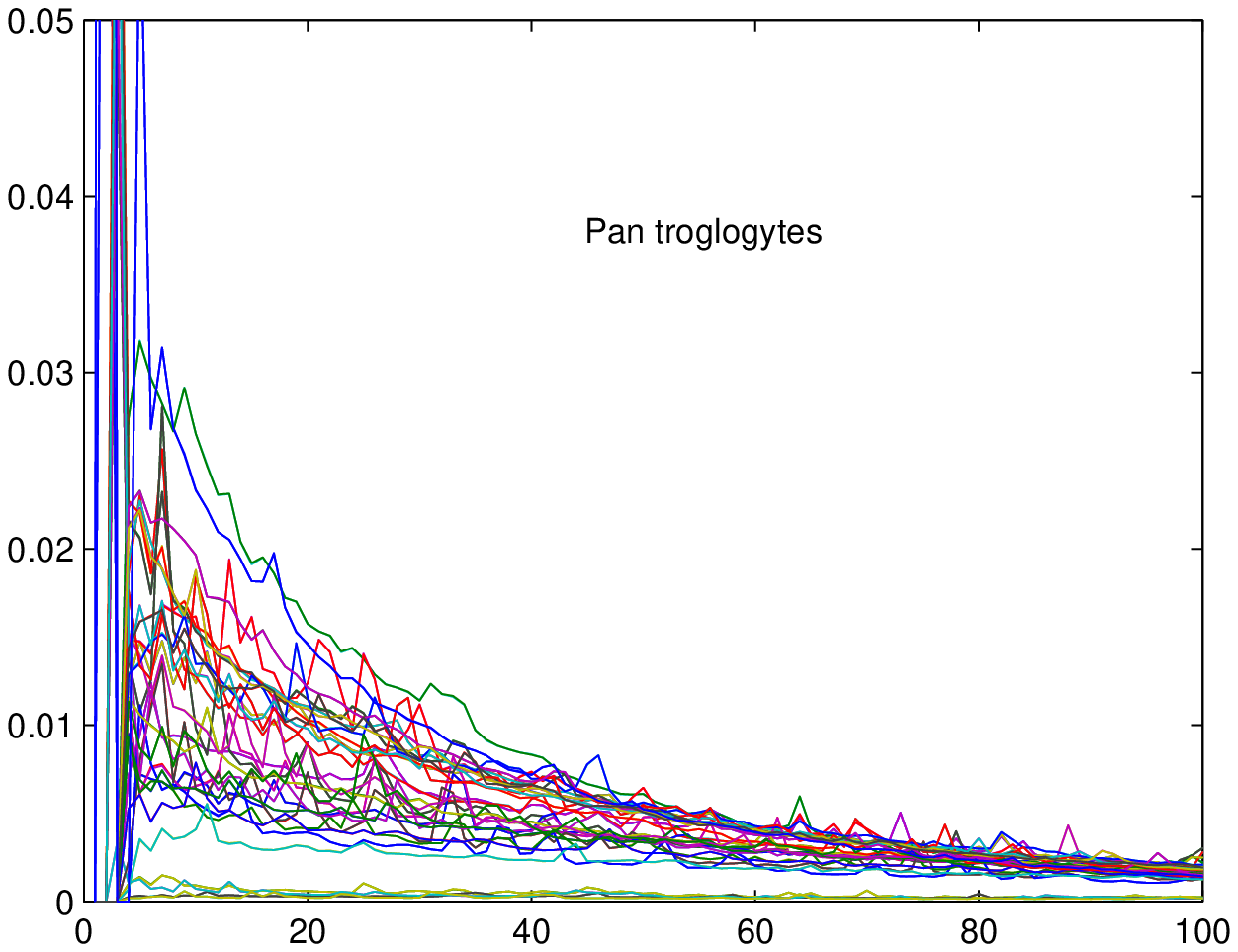}
 {\small \bf b} \includegraphics[width=8.5cm]{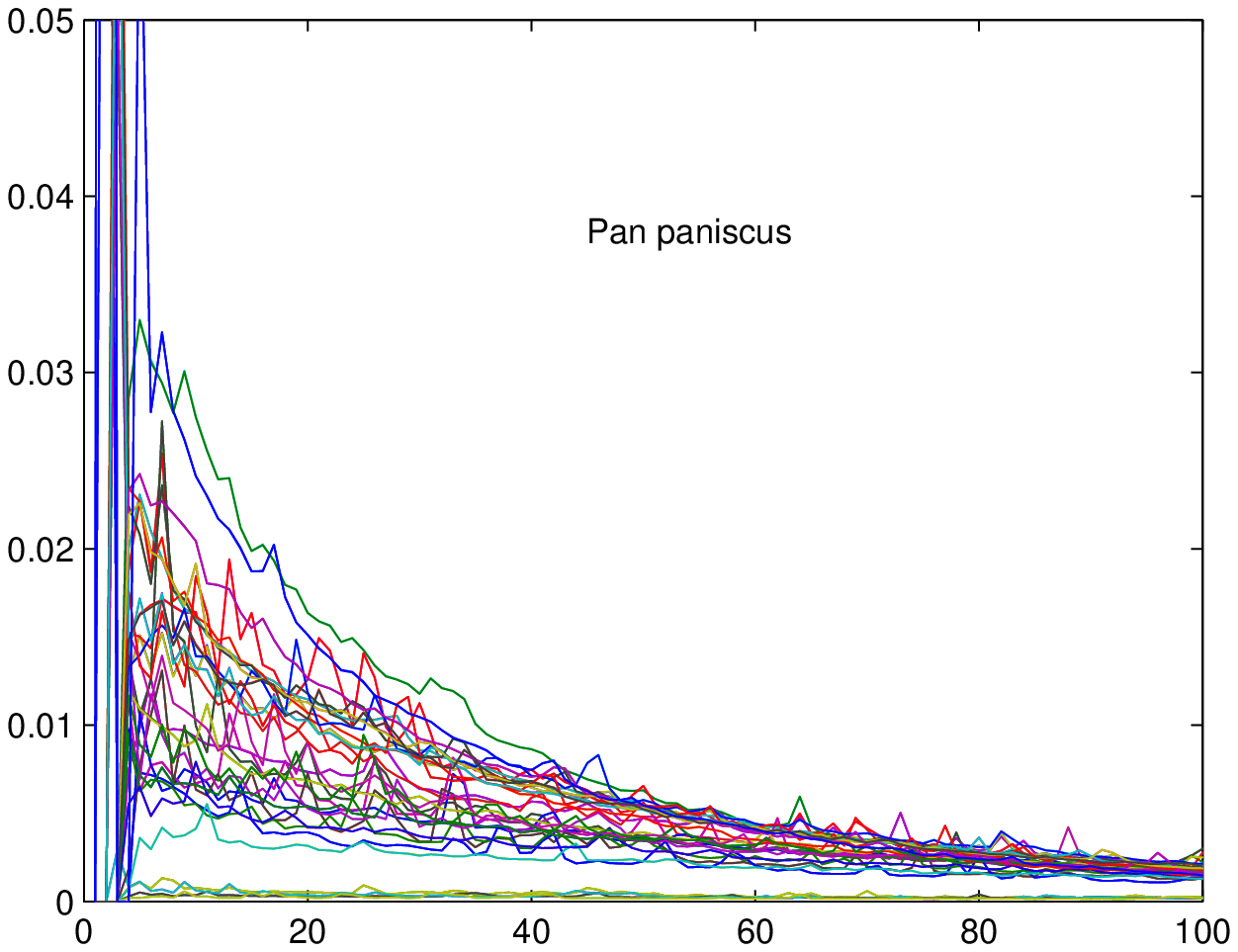}\\
 {\small \bf c} \includegraphics[width=8.5cm]{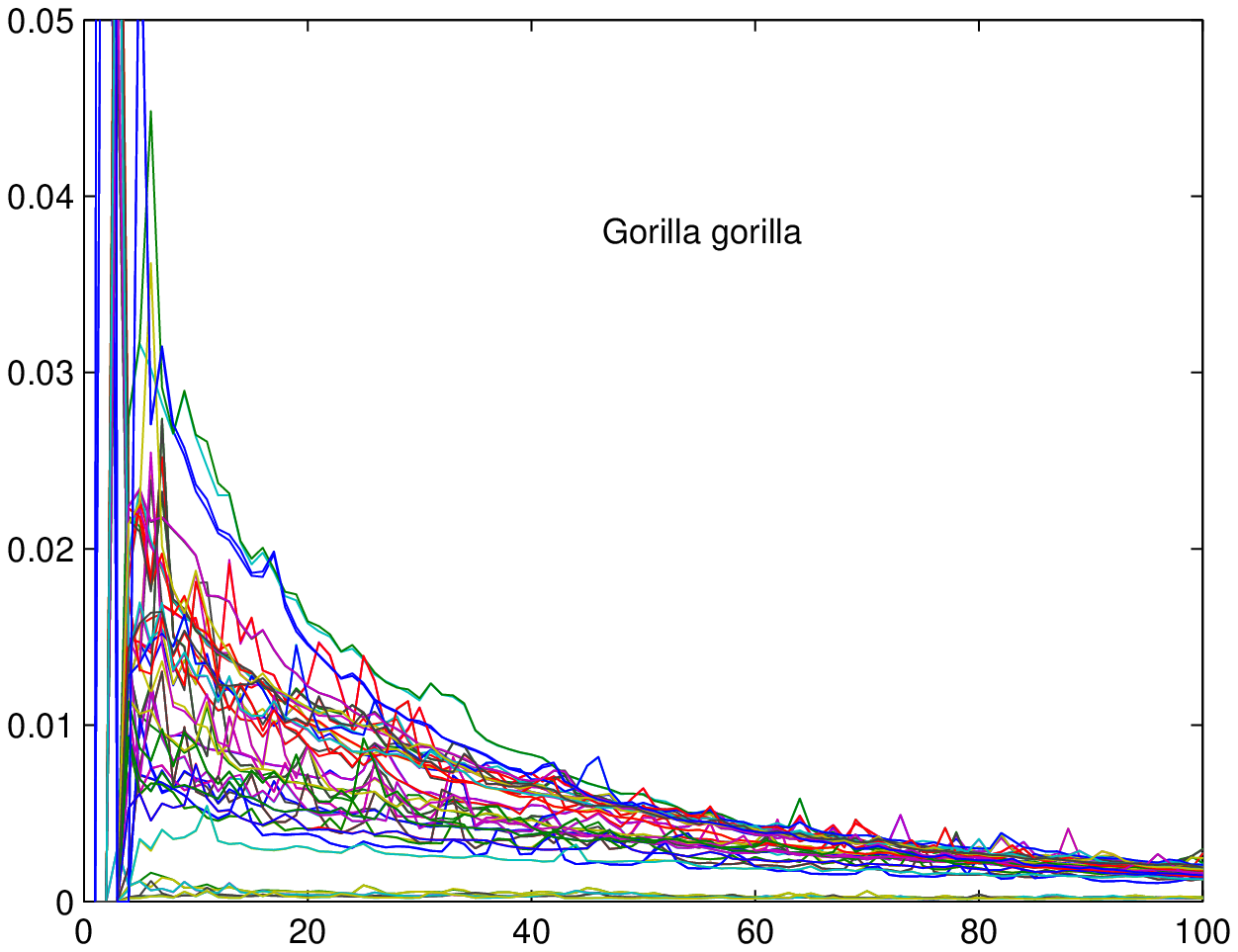}
 {\small \bf d} \includegraphics[width=8.5cm]{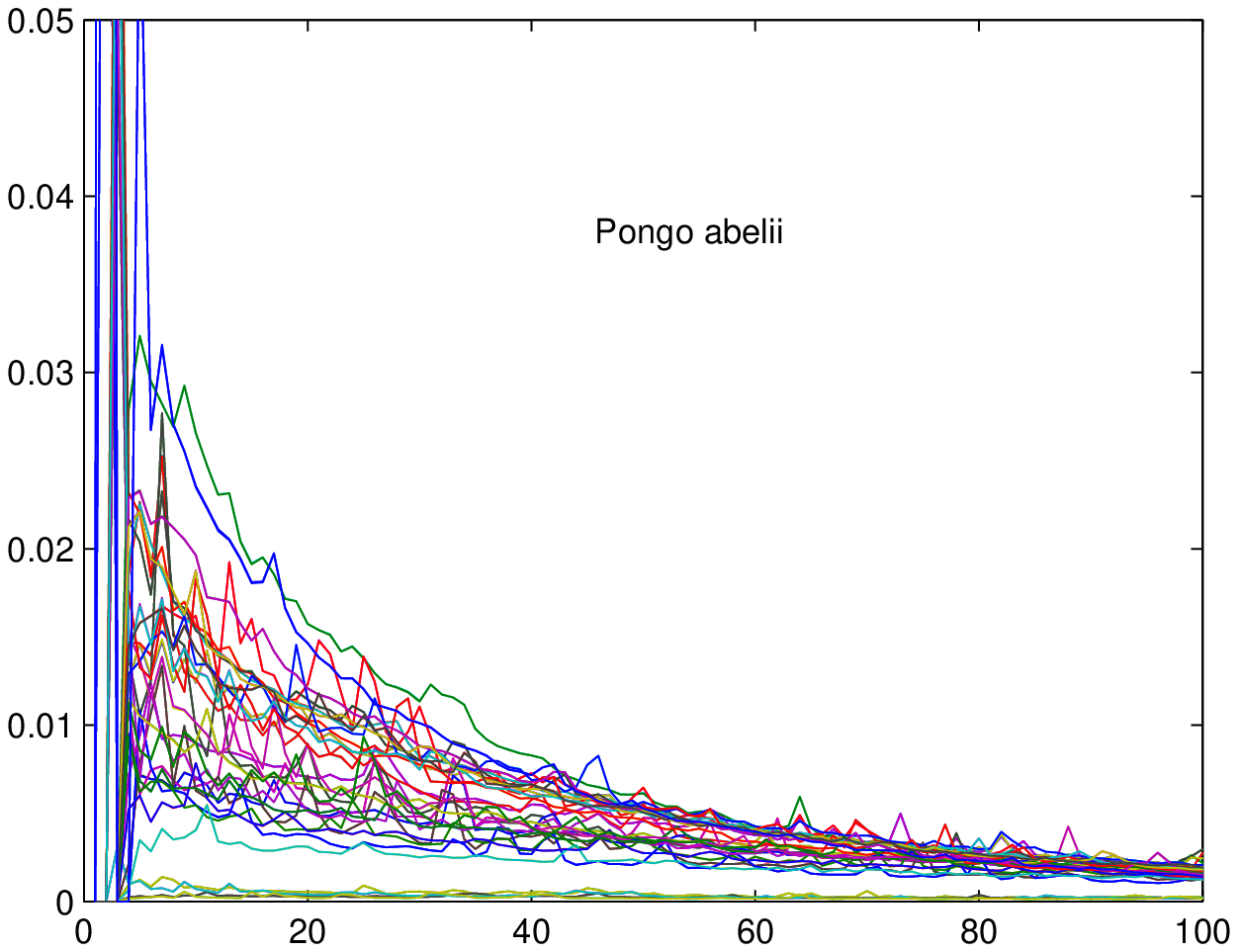}
\end{figure}

\clearpage  
\begin{figure}
 \centering
 {\small \bf e} \includegraphics[width=8.5cm]{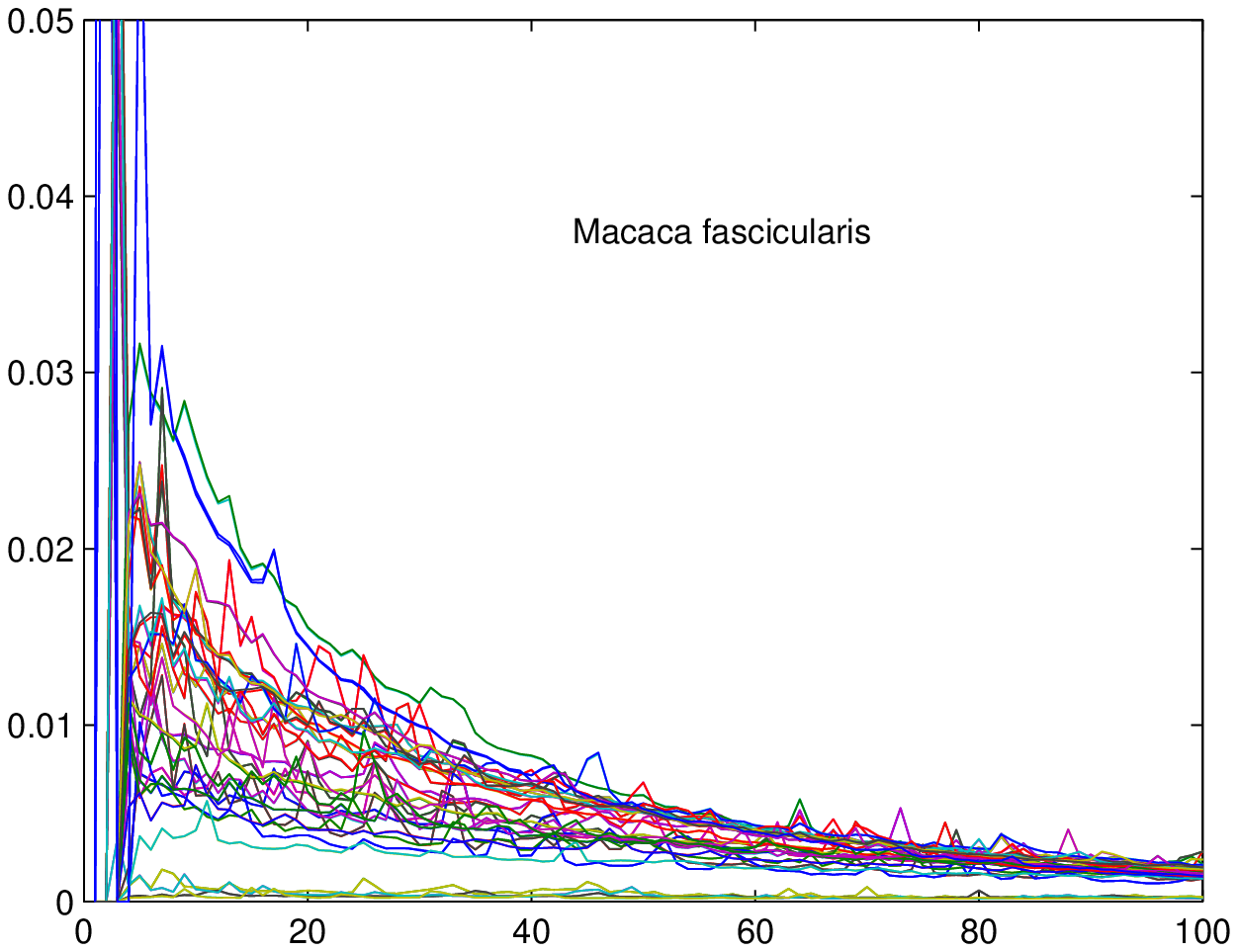}
 {\small \bf f} \includegraphics[width=8.5cm]{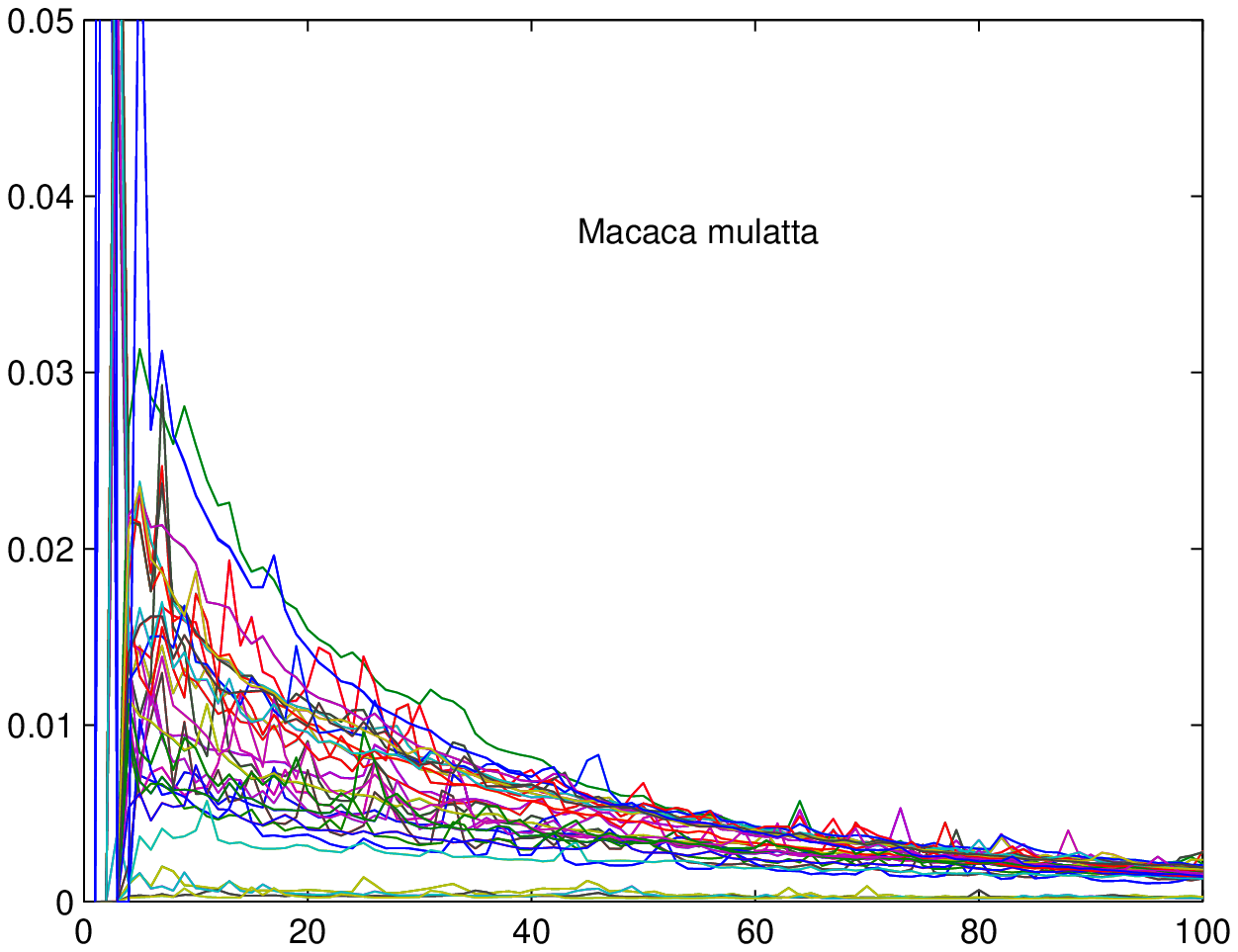}\\
 {\small \bf g} \includegraphics[width=8.5cm]{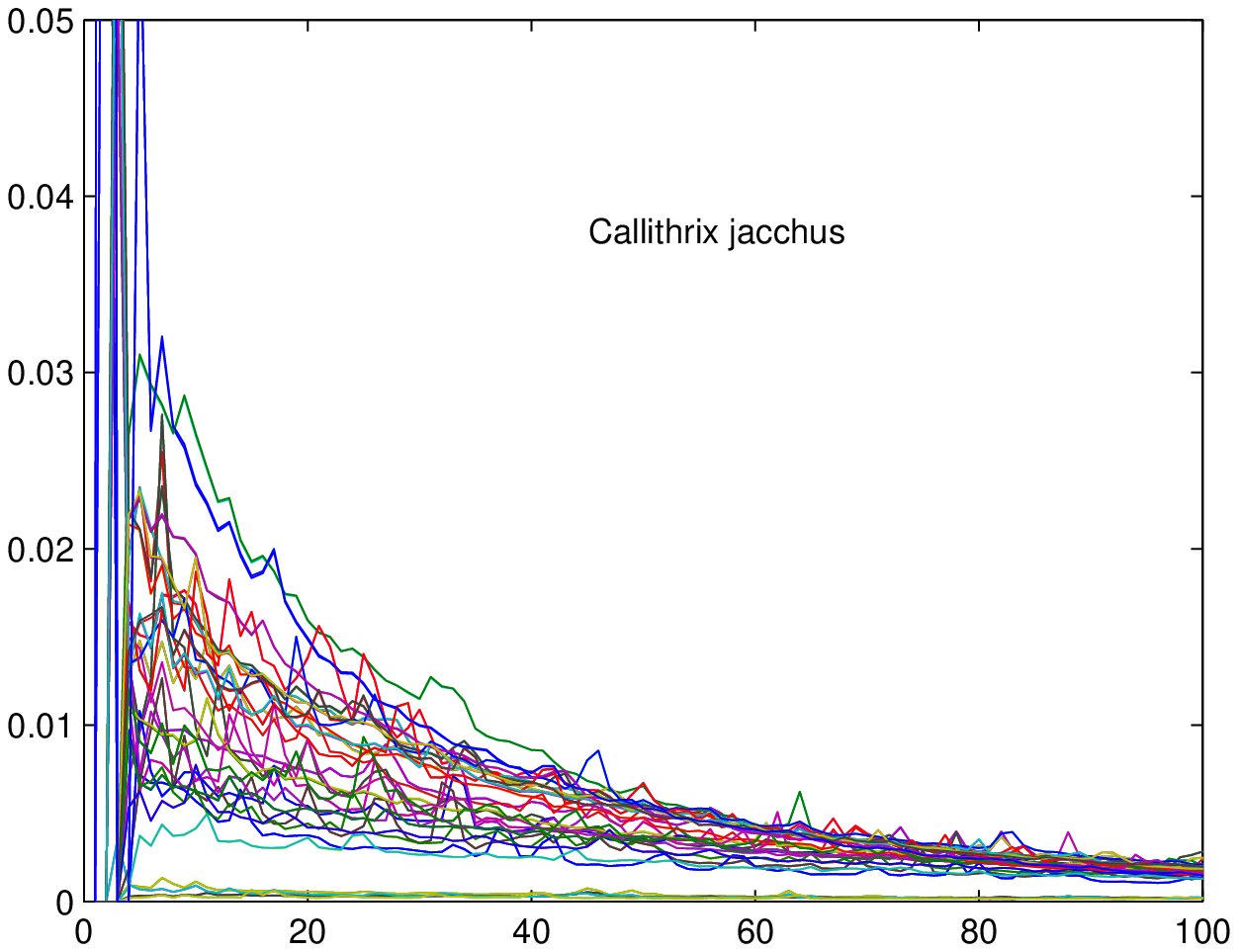}
 {\small \bf h} \includegraphics[width=8.5cm]{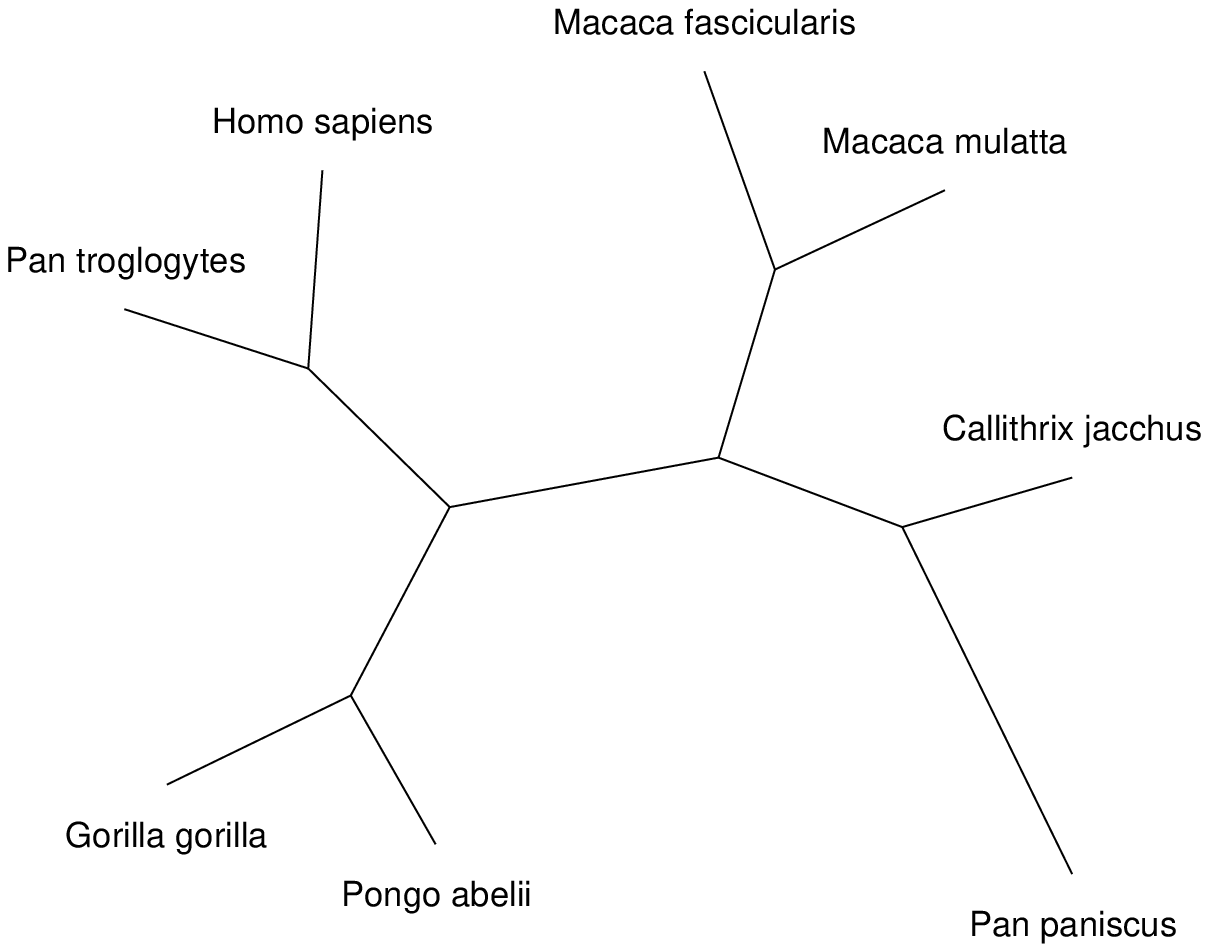}
 \caption{The tree of primates obtained by comparing their genomic codon distributions. {\bf a} Pan troglodytes. {\bf b} Pan paniscus. {\bf c} Gorilla gorilla. {\bf d} Pongo abelii. {\bf e} Macaca fascicularis. {\bf f} Macaca mulatta. {\bf g} Callithrix jacchus. {\bf h} Tree of primates.}
\end{figure}

\clearpage  
\begin{figure}
 \centering
 {\small \bf a} \includegraphics[width=8.5cm]{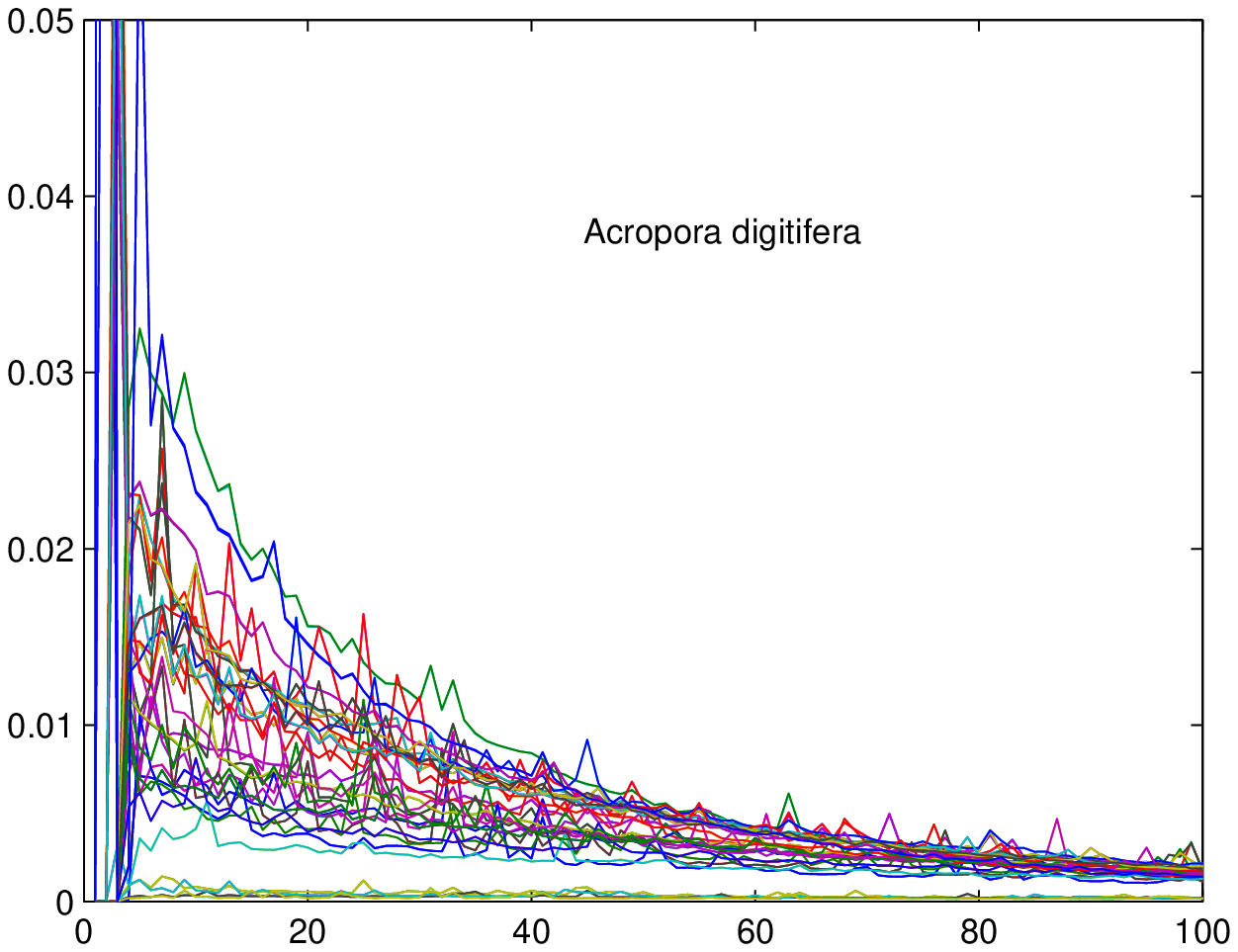}
 {\small \bf b} \includegraphics[width=8.5cm]{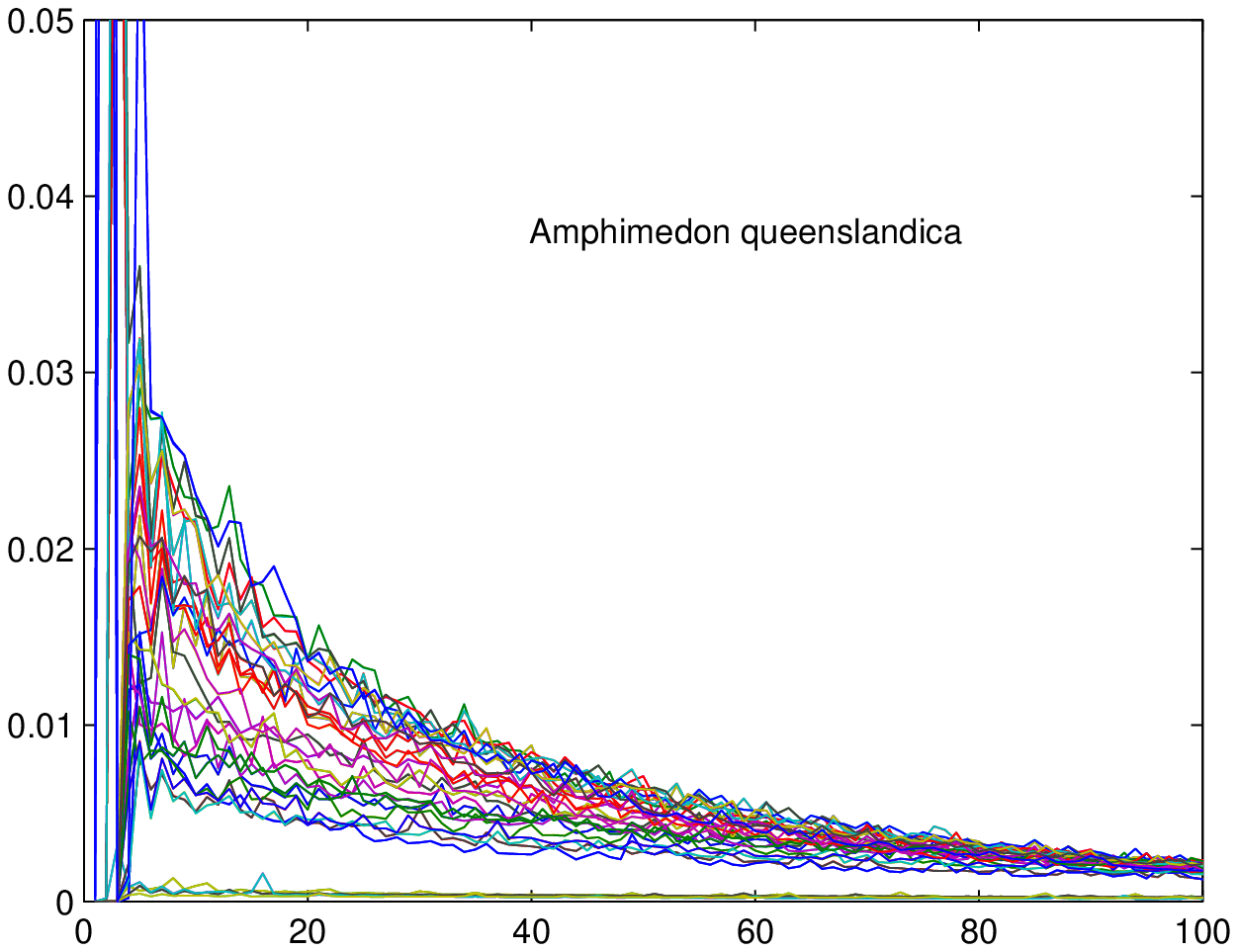}\\
 {\small \bf c} \includegraphics[width=8.5cm]{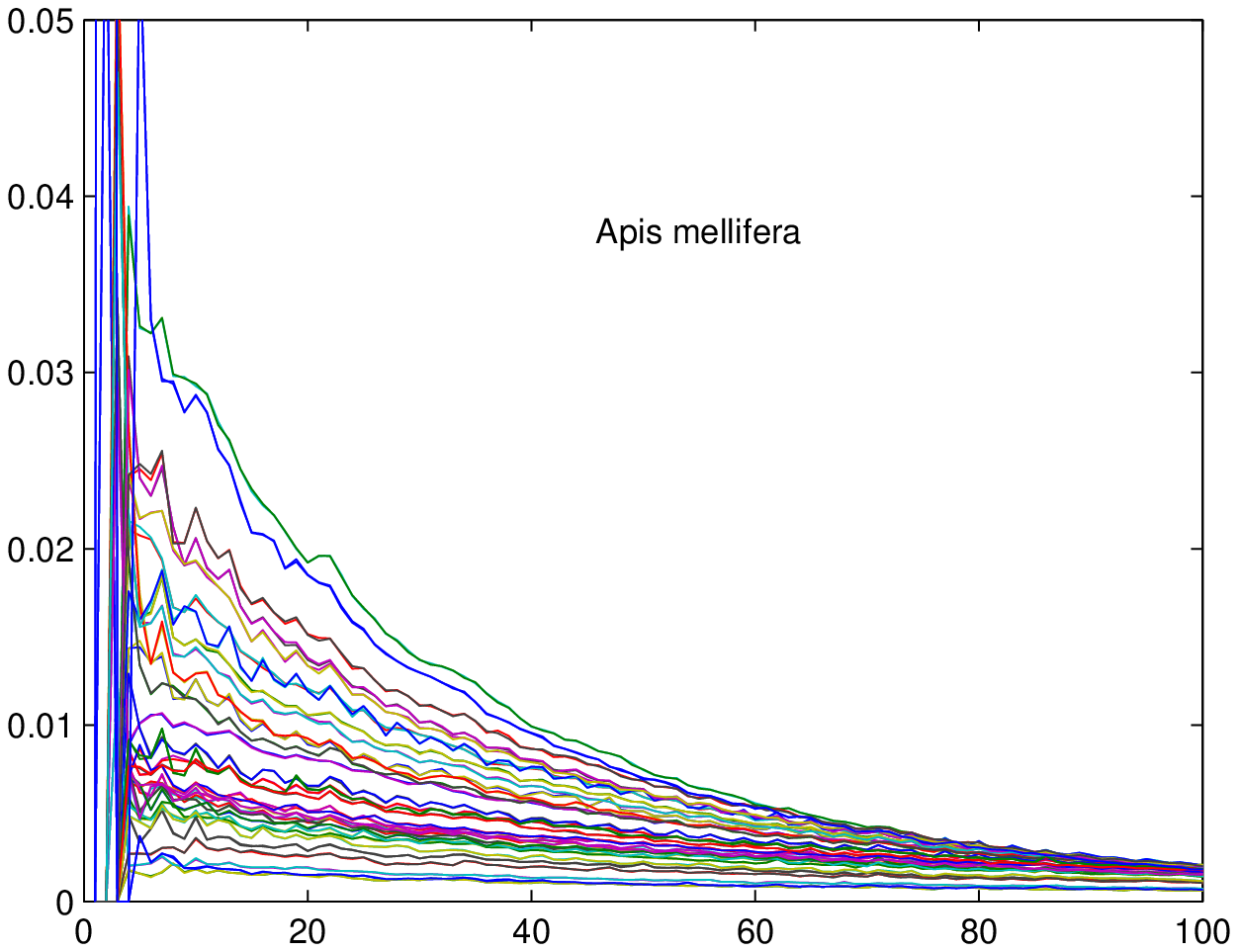}
 {\small \bf d} \includegraphics[width=8.5cm]{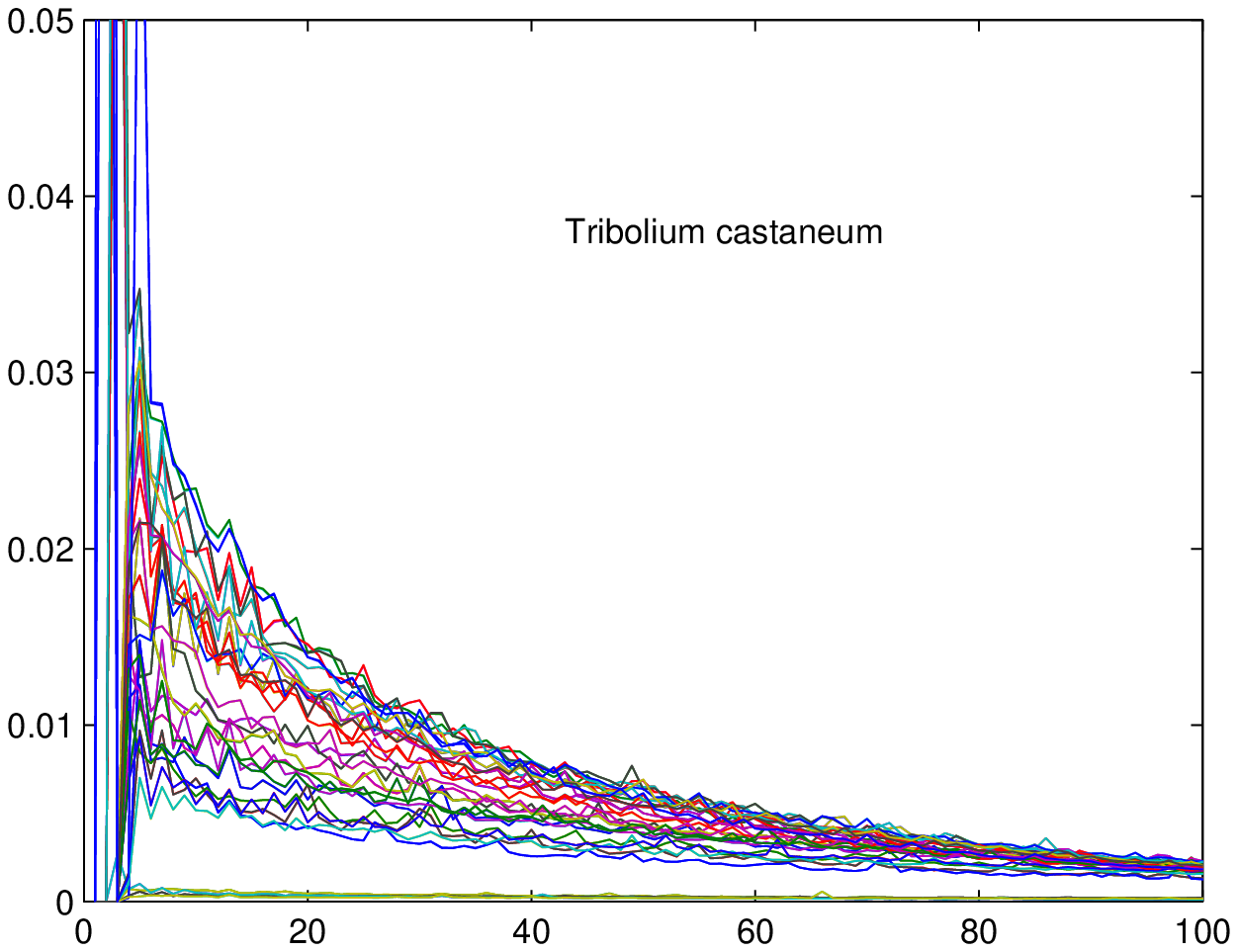}\\
 {\small \bf e} \includegraphics[width=8.5cm]{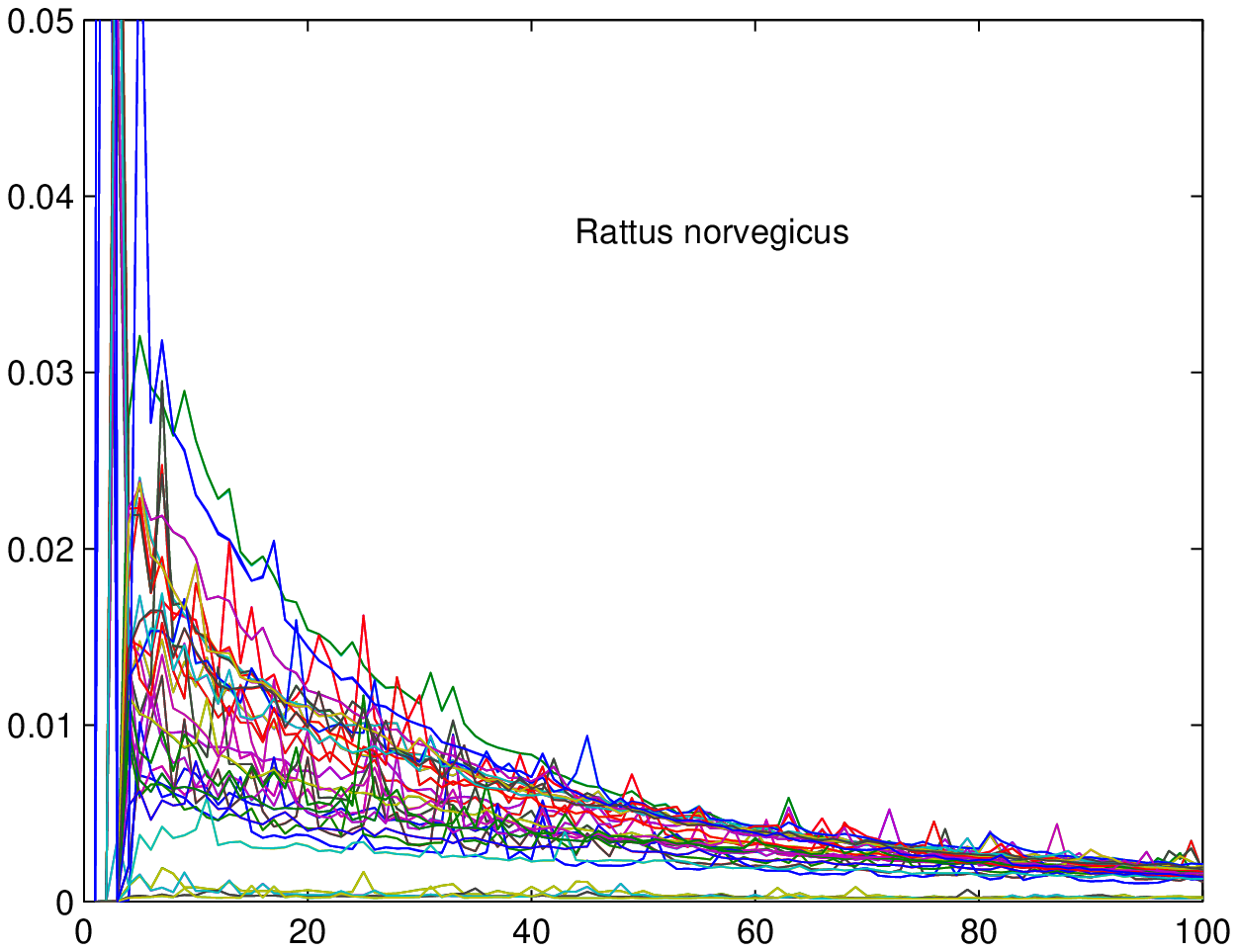}
 {\small \bf f} \includegraphics[width=8.5cm]{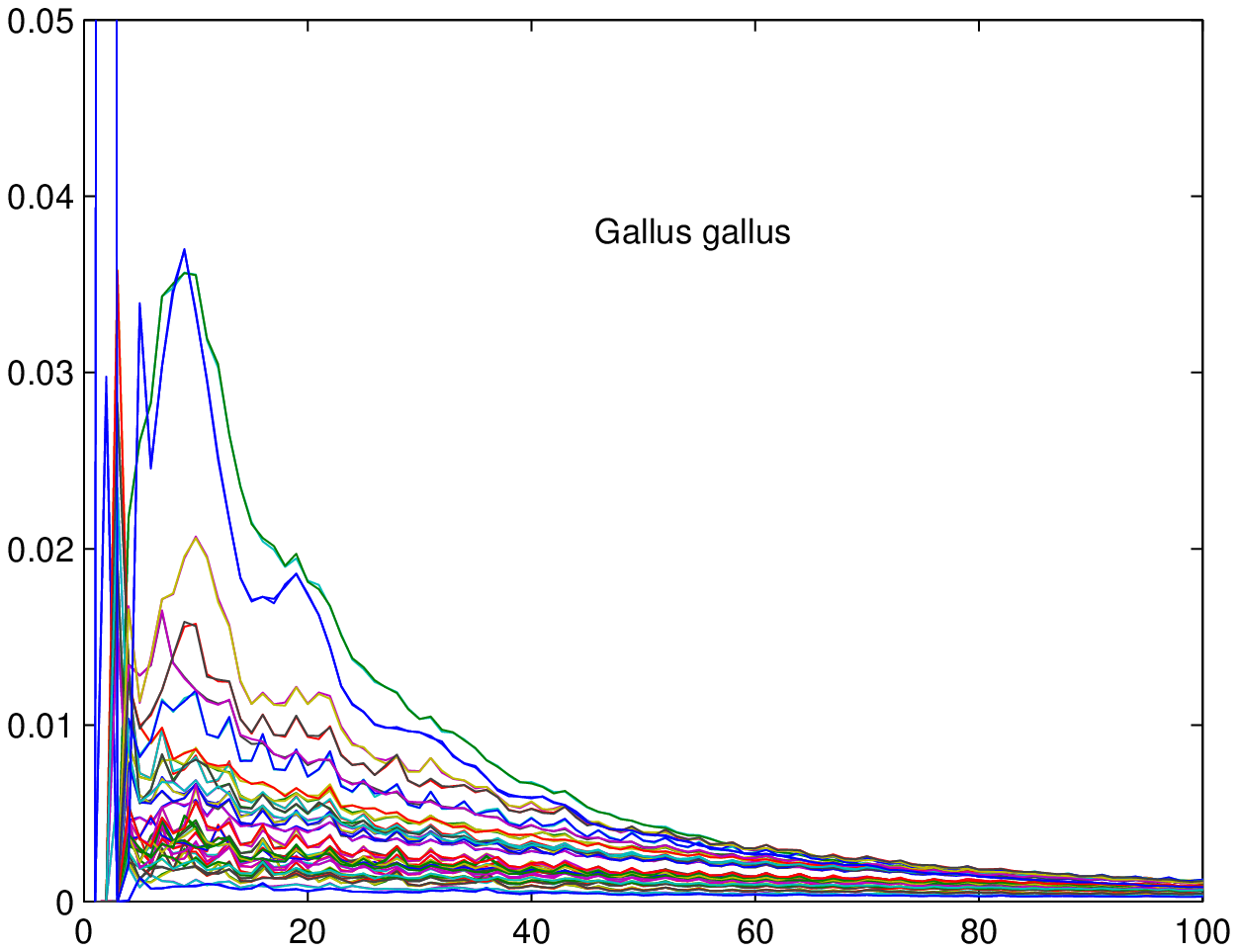}
 \caption{Comparing the genomic codon distributions for animals in the three subkingdoms. Diploblastica: {\bf a} Acropora digitifera, {\bf b} Amphimedon queenslandica; Protostomia: {\bf c} Apis mellifera, {\bf d} Tribolium castaneum; Deuterostomia: {\bf e} Rattus norvegicus, {\bf f} Gallus gallus.}
\end{figure}

\clearpage  
\begin{figure}
 \centering
 \includegraphics[width=18.3cm]{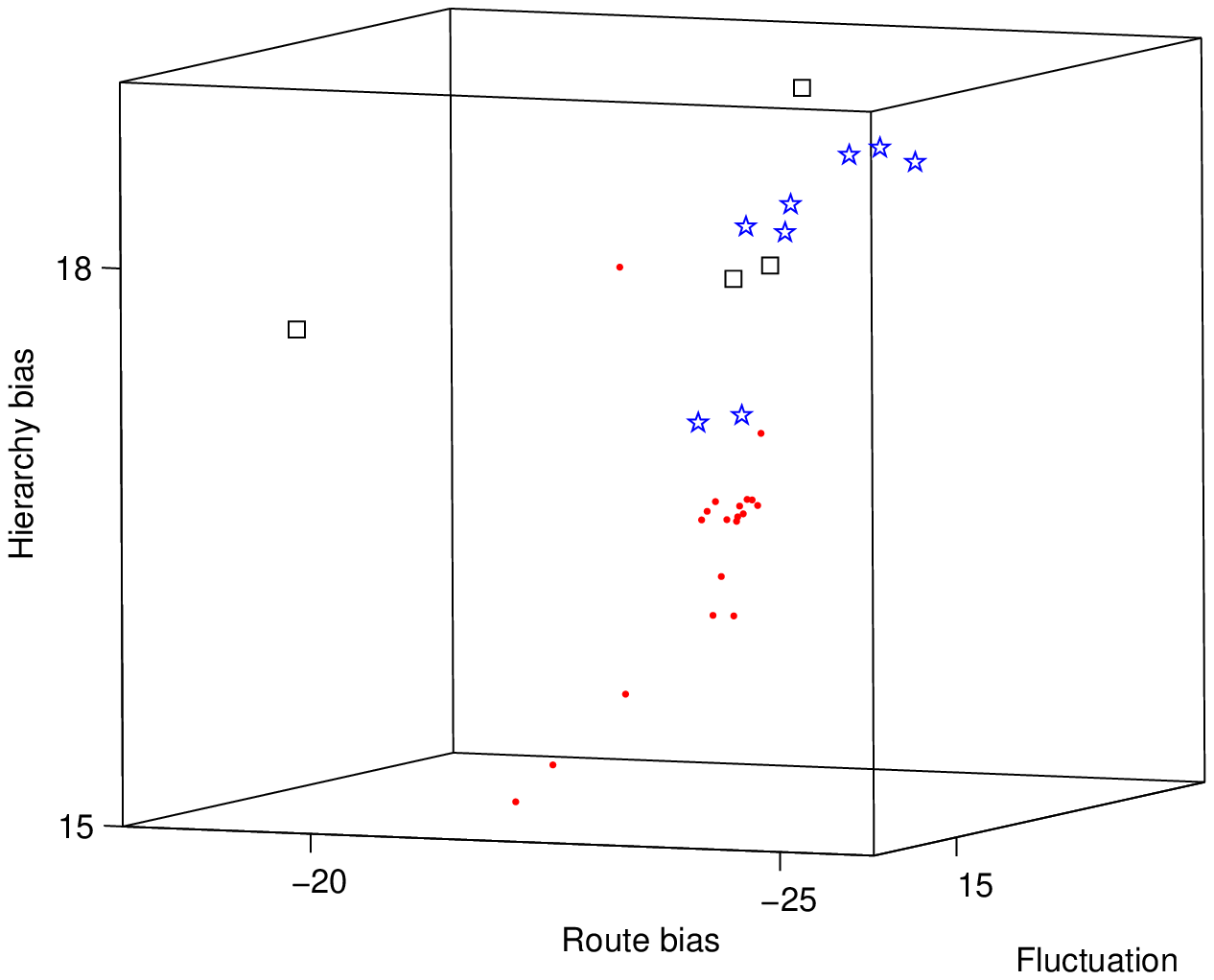}
 \caption{Separate clusterings of species from the three subkingdoms of animals in the biodiversity space may support the separate origins of the three subkingdoms during the Cambrian explosion proposed as a hypothesis based on fossil records. The squares represent Diploblastica (Acropora digitifera, Amphimedon queenslandica, Hydra magnipapillata, Stylophora pistillata), the stars represent Protostomia (Apis mellifera, Caenorhabditis elegans, Tribolium castaneum, Aedes aegypti, Crassostrea virginica, Pomacea canaliculata, Tribolium castaneum, Xenopus laevis), and the dots represent Deuterostomia (Bos taurus, Canis familiaris, Danio rerio, Equus caballus, Gallus gallus, Monodelphis domestica, Mus musculus, Rattus norvegicus and the above primates).}
\end{figure}

\end{document}